\documentclass[aps,prb,twocolumn,superscriptaddress,amsmath]{revtex4-1}
\usepackage[caption=false]{subfig}
\usepackage{graphicx,bm,wasysym,color,multirow}
\usepackage{amsmath,amsfonts,amsthm,amssymb,amsbsy,bbold,mathrsfs}
\definecolor{orange}{rgb}{1,0.5,0}

\begin{document}
\newcommand{\comment}[1]{\color{red} #1 \color{black}}
\newcommand{\sidecomment}[1]{\marginpar{\color{red} #1 \color{black}}}
\newcommand{\h}{$\mathcal{H}$-}
\newcommand{\shc}{stripyhoneycomb }
\newcommand{\hh}{hyperhoneycomb }
\def \li213{Li$_2$IrO$_3$}
\def \na213{Na$_2$IrO$_3$}
\newcommand{\ort}[1]{\frac{#1}{\sqrt{2}}}
\newcommand{\bu}{\bar{u}}
\newcommand{\angstrom}{\text{\normalfont\AA}}

\title{Three dimensional quantum spin liquids in models of harmonic-honeycomb iridates 
and phase diagram in an infinite-D approximation}
\author{Itamar Kimchi}
\affiliation{Department of Physics, University of California, Berkeley, CA 94720, USA}
\author{James G. Analytis}
\affiliation{Department of Physics, University of California, Berkeley, CA 94720, USA}
\affiliation{Materials Science Division, Lawrence Berkeley National Laboratories, Berkeley, CA 94720, USA}
\author{Ashvin Vishwanath}
\affiliation{Department of Physics, University of California, Berkeley, CA 94720, USA}
\affiliation{Materials Science Division, Lawrence Berkeley National Laboratories, Berkeley, CA 94720, USA}
\begin{abstract}
Motivated by the recent synthesis of two insulating Li$_2$IrO$_3$ polymorphs, where Ir$^{4+}$ $S_{\rm eff}{=}1/2$ moments form 3D (``harmonic'') honeycomb structures with threefold coordination, we study magnetic Hamiltonians on the resulting $\beta$-Li$_2$IrO$_3$ hyperhoneycomb lattice and $\gamma$-Li$_2$IrO$_3$ stripyhoneycomb lattice. Experimentally measured  magnetic susceptibilities suggest that Kitaev interactions, predicted for the ideal 90$^\circ$ Ir-O-Ir bonds, are sizable in these materials.  We first consider pure Kitaev interactions,  which lead to an exactly soluble 3D quantum spin liquid (QSL) with emergent Majorana fermions and Z$_2$~flux loops. Unlike 2D QSLs, the 3D QSL  is stable to finite temperature, with $T_c  \approx |K|/100$.   On including Heisenberg couplings, exact solubility is lost. However, by noting that the shortest closed loop $\ell$ is relatively large in these structures, we construct an $\ell\rightarrow \infty$ approximation by defining the model on the Bethe lattice. The phase diagram of the Kitaev-Heisenberg model on this lattice is obtained directly in the thermodynamic limit, using tensor network states and the infinite-system time-evolving-block-decimation (iTEBD) algorithm. Both magnetically ordered and gapped QSL phases are found, the latter being identified by an entanglement fingerprint. 
\end{abstract}
\maketitle
\section{Introduction}
Recently there has been growing interest in studying quantum phases of matter that are characterized by long range entanglement\cite{WenBook}, in contrast to conventional symmetry broken states. 
In particular, gapped quantum phases that feature long range entanglement exhibit remarkable emergent properties such as excitations with unusual statistics and fractional quantum numbers. These properties are known to occur in two dimensional phases such as the fractional quantum Hall states, which are realized in 2D electron gases in strong magnetic fields. In solids, frustrated insulating magnets are believed to be prime candidates for avoiding conventional ordering in favor of a long range entangled phase of matter --- the quantum spin liquid phase. Recent numerical studies have found mounting evidence for gapped spin liquids, phases which are long range entangled\cite{Sachdev1992,Vishwanath2006,Lee2011,Motrunich2011,Balents2012,Schollwock2012}, on two dimensional geometrically frustrated lattices such as the Kagome lattice\cite{Elser1993,Misguich2000,White2011}. 

\begin{figure}[b]
\includegraphics[width=\columnwidth]{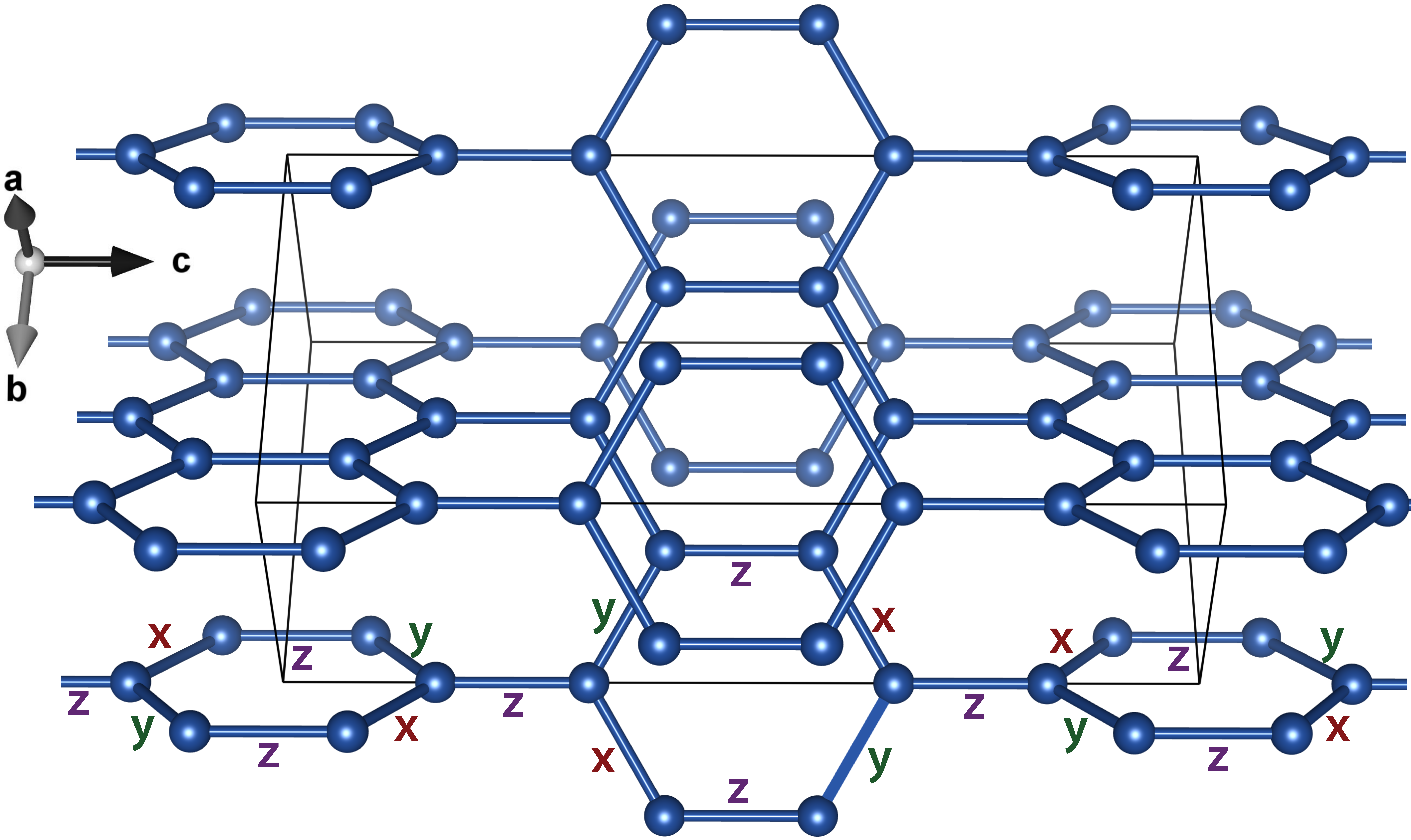}
\caption[]{{\bf The \shc lattice of iridium in \hspace{1em}	$\gamma$-Li$_2$IrO$_3$. } The recently synthesized \shc lattice (space group \#66 \textit{Cccm}) has threefold coordinated sites, which form hexagons arranged in stripes of alternating orientation. 
It is the $n{=}1$ member of the harmonic honeycomb\cite{analytis2014} series of structures; the distinct \hh lattice of $\beta$-\li213 (Fig.~\ref{fig:hhc0}) has $n{=}0$. 
Parent orthorhombic coordinate system and unit cell (boxed) are shown. 
In the limit of superexchange via ideal oxygen octahedra, the magnetic Hamiltonian is dominated by Kitaev-type couplings (\textsf{x,y,z} labels at bottom), leading to an exactly solvable model of a 3D quantum spin liquid. }
\label{fig:hhc1}
\end{figure}

However,  frustration need not arise from geometry alone.   In quantum magnets of heavy elements, spin-orbit coupling leads to anisotropic interactions that may engender quantum disordered ground states even in the absence of the usual geometrical frustration. A prime example is the honeycomb lattice -- a bipartite lattice on which both ferromagnetic and antiferromagnetic Heisenberg couplings host ordered ground states. However, a peculiar set of anisotropic interactions proposed by Kitaev\cite{Kitaev2006}, where  neighboring spins are coupled by Ising interactions along an axis that is set by the spatial orientation of the bond, has been shown to be in a quantum spin liquid phase. Furthermore, this is demonstrated via an exact solution -- in contrast to the numerical tour de force required for identifying the spin liquid phase in the Kagome antiferromagnet\cite{White2011,Balents2012}.  

Interestingly, the requirement for obtaining an exactly soluble spin liquid is not specific to the honeycomb lattice. Instead, the key ingredients are the three fold coordination of the sites and the peculiar Ising interaction with rotating axes. If such a network would be created in three dimensions, it would lead to an example of a 3D quantum spin liquid. Such long range entangled quantum phases in 3D are less well explored than their 2D counterparts. While basic constraints on long range entangled quantum phases in 3D have been discussed\cite{LevinWen,Vishwanath2011}, few suggestions for materials candidates exist.   An exception is the 3D hyperkagome material\cite{Takagi2007}  Na$_4$Ir$_3$O$_8$, for which a spin liquid ground state with bosonic\cite{Vishwanath2008} or fermionic\cite{Lee2008,Paramekanti} spinon excitations has been proposed.  Related U(1) spin liquids\cite{Hermele,Sondhi} have been proposed for quantum spin ice materials\cite{BalentsSpinice} on the pyrochlore lattice. Here we discuss a 3D example of quantum spin liquid behavior induced by spin-orbit coupling in a 3D model with Kitaev exchanges, and explore a possible physical realization. 

\begin{figure}[t]
\includegraphics[width=220 pt]{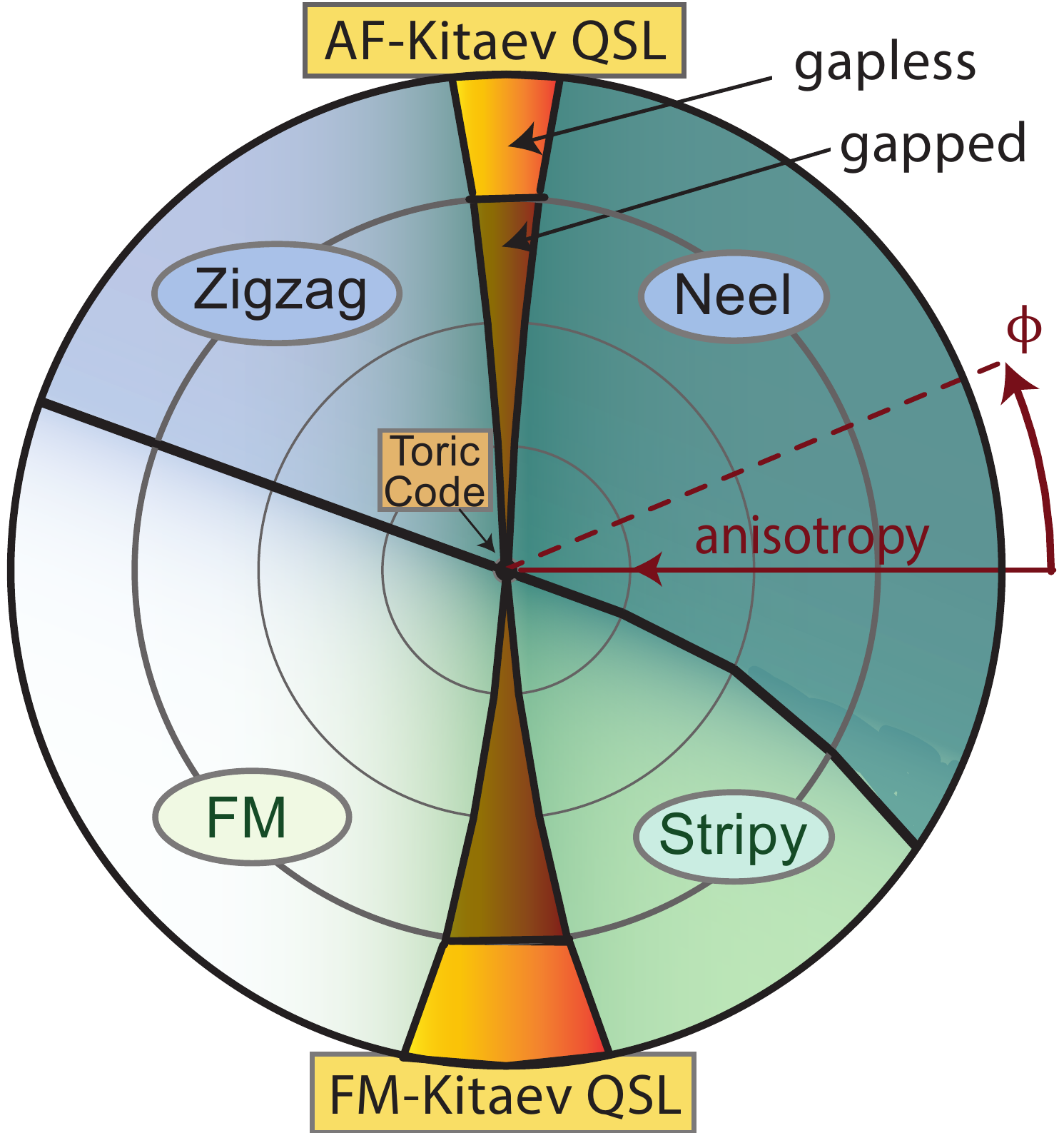}
\caption[]{{\bf Quantum phase diagram in the large-$\ell$ limit. }
Phase diagram of the frustrated quantum Hamiltonian Eq.~\ref{eq:KH1}, computed via tensor network states within an infinite-D or large-$\ell \rightarrow \infty$ approximation to the hyperhoneycomb's $\ell=10$. Except for quantitative extent of QSLs (not to scale), we expect it to describe the \shc and hyperhoneycomb lattices of $\gamma$- and $\beta$-\li213, for which we argue this is a physical model. 
The 2-parameter space shown here (Eqs.~\ref{eq:KH1},\ref{eq:KH2}) has polar axes 
$r$ tuning symmetry-allowed Kitaev bond anisotropy and $\phi$ setting relative strength of Kitaev and Heisenberg  interactions. 
The QSL phases, successfully stabilized on the Bethe lattice by the algorithm's finite entanglement cutoff $\chi$, were identified by an entanglement fingerprint. 
}
\label{fig:radial_phases}
\end{figure}

\begin{figure}
\includegraphics[width=\columnwidth]{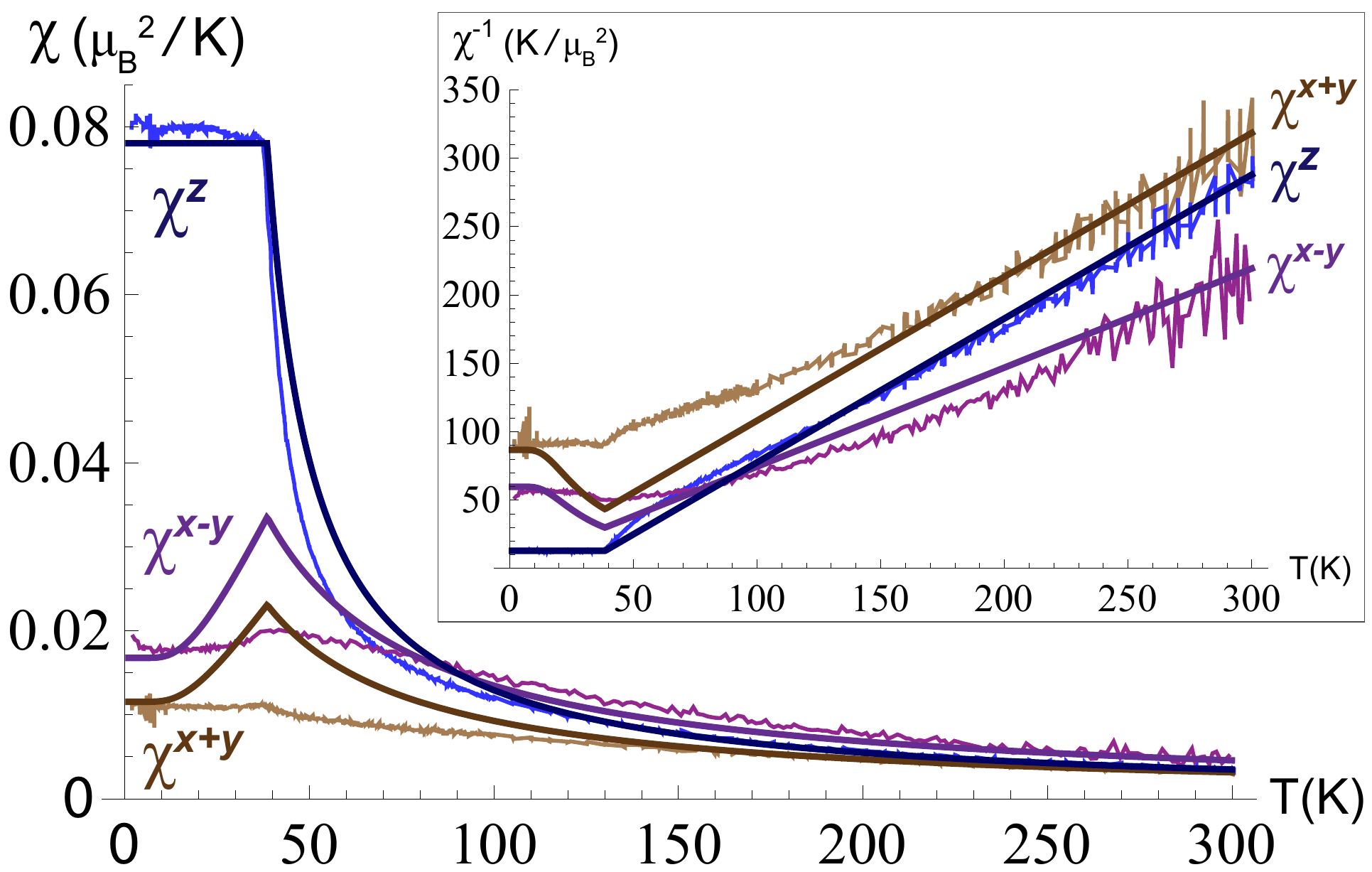}
\caption[]{{\bf Strong Kitaev exchanges capturing $\gamma$-\li213 anisotropic susceptibility. } 
Magnetic susceptibility (inset: inverse susceptibility) along principal axes
$(z{=}b;x{\pm}y{=}a,c)$, 
measured\cite{analytis2014} for a $\gamma$-\li213 crystal (bright lines) and theoretical mean field fit (dark lines). 
Susceptibility is fitted by the minimal Hamiltonian Eq.~\ref{eq:KH1} with parameters $(K_c,K_d,J_c,J_d)$ at $(-17,-7,6.3,0.8)$ meV; magnetic order (recently found\cite{RaduSpiral} to be a noncoplanar spiral) is captured\cite{RaduSpiral} by Eq.~\ref{eq:KH1} at  $(-15,-12,5,2.5)$ meV, supplemented by $c$-axis Ising exchange on $c$-bonds and $J_2$ Heisenberg exchange on second-neighbors. 
In both cases, large   Kitaev exchanges $K_c,K_d$  are necessary to describe the material.
}
\label{fig:fits}
\end{figure}

At first sight, the Kitaev interactions seem rather unphysical. However, as pointed out by Jackeli and Khaliullin\cite{Khaliullin2009}, they may actually be realized under certain circumstances in iridium oxides. An Ir$^{4+}$ ion at the center of an oxygen octahedron is expected to be in a Kramers doublet state $J{=}1/2$, with the doublet wave function set by the spin-orbit coupling. This leads to unusual magnetic exchange interactions. For example, when a pair of Ir$^{4+}$  moments are coupled via an intermediate oxygen with a 90$^\circ$ bond, the usual Goodenough-Kanamori-Anderson rules would have predicted a ferromagnetic Heisenberg exchange. Here however, due to the special nature of the Kramers doublets, the coupling was shown\cite{Khaliullin2009} to be ferromagnetic, but of the Ising type, with the spin component involved being perpendicular to the bond's iridium-oxygen plane. 
Other exchange paths around the Ir-O-Ir-O square and involving higher energy states including the Ir$^{4+}$ $e_g$ orbitals\cite{Khaliullin2010,Khaliullin2012} also generate this type of coupling, with either sign. 
For the compound Na$_2$IrO$_3$ in which Ir forms independent honeycomb lattices, these mechanisms were argued\cite{Khaliullin2009} to lead to couplings identical to Kitaev's honeycomb model, although additional spin interaction, minimally a Heisenberg term, is also expected.
An appropriate minimal model for the low energy magnetic Hamiltonian is then the nearest neighbor Kitaev-Heisenberg model\cite{Khaliullin2010}. 

In the \textit{C2/m} layered structure\cite{Taylor2012} of Na$_2$IrO$_3$, and even more dramatically in the \textit{Cccm} and  \textit{Fddd} 3D-\li213 structures we discuss below\cite{analytis2014}, space group symmetries single out the subset of Ir-Ir bonds which are oriented along a particular axis. Recent ab initio work\cite{Imada2014} has found that already for Na$_2$IrO$_3$, the magnitude of both Kitaev and Heisenberg couplings can be quite different between these symmetry-distinguished subsets of bonds.  Allowing the couplings to take a different value on the symmetry-distinguished ``$c$-bonds'' compared to the remaining ``$d$-bonds'' produces the bond-anisotropic Kitaev-Heisenberg Hamiltonian,
\begin{align}
\label{eq:KH1}
H &=\sum_{c-\text{bond} \langle ij\rangle} \left( K_c S_i^{\gamma_{ij}} S_j^{\gamma_{ij}} + J_c \vec{S_i} \cdot \vec{S_j} \right) \\
 &+\sum_{d-\text{bond} \langle ij\rangle} \left( K_d S_i^{\gamma_{ij}} S_j^{\gamma_{ij}} + J_d \vec{S_i} \cdot \vec{S_j} \right) . \nonumber
\end{align}
The geometry of IrO$_6$ octahedra implies that the spin component $\gamma_{ij}$ coupled in a Kitaev term $K S_i^{\gamma_{ij}} S_j^{\gamma_{ij}}$ is, on any bond, one of the Ir-O Cartesian axes $x$, $y$ or $z$.

The additional Heisenberg interactions are important; indeed, the ground state of Na$_2$IrO$_3$ is magnetically ordered and not a quantum spin liquid. 
The ``zigzag'' (wavevector $M$) magnetic ordering seen\cite{Hill2011,Taylor2012,Cao2012} in Na$_2$IrO$_3$, as well as other measured magnetic and electronic properties, remain consistent with Kitaev-Heisenberg as well as with more conventional Hamiltonians with SU(2) rotation symmetry.\cite{Gegenwart2010,You2011,Lauchli2011,Gegenwart2012,Taylor2012,Khaliullin2012,Damascelli2012,Valenti2013,Min2013,Kim2013a,Kim2013}
Other anisotropic exchanges related to the Jackeli-Khaliullin mechanism\cite{Khaliullin2009} have been described\cite{Khaliullin2005,Balents2008,Micklitz2010,Norman2010} for Na$_2$IrO$_3$ and related iridates\cite{Kaul2013,Gegenwart2014,Gegenwart2014a}.
Alternative starting scenarios for Na$_2$IrO$_3$ have also been proposed\cite{Nagaosa2009,Khomskii2012,Kim2012a,Valenti2013a} which paint a picture of it different from a Mott insulator.  
Since the Chaloupka et al original formulation and solution of the Kitaev-Heisenberg model\cite{Khaliullin2010}, much research has elucidated its various properties\cite{Gegenwart2012,Trebst2011,Trebst2011b,Rachel2012,Daghofer2012,KHbhc}; as a model containing a QSL, it has been especially interesting to investigate its behavior under charge doping\cite{Vishwanath2012,Rosenow2012,Okamoto2013,Horsch2013,You2013}. 
While the Kitaev-Heisenberg model may or may not apply to the particular compound Na$_2$IrO$_3$, the key point is that the Jackeli-Khaliullin mechanism can arise in any lattice of edge-sharing IrO$_6$ octahedra with roughly cubic local symmetry, as long as any distortion from cubic symmetry is weaker than the spin orbit coupling\cite{Khaliullin2009,Khaliullin2012,KHbhc}. 

\begin{figure}
\includegraphics[width=\columnwidth]{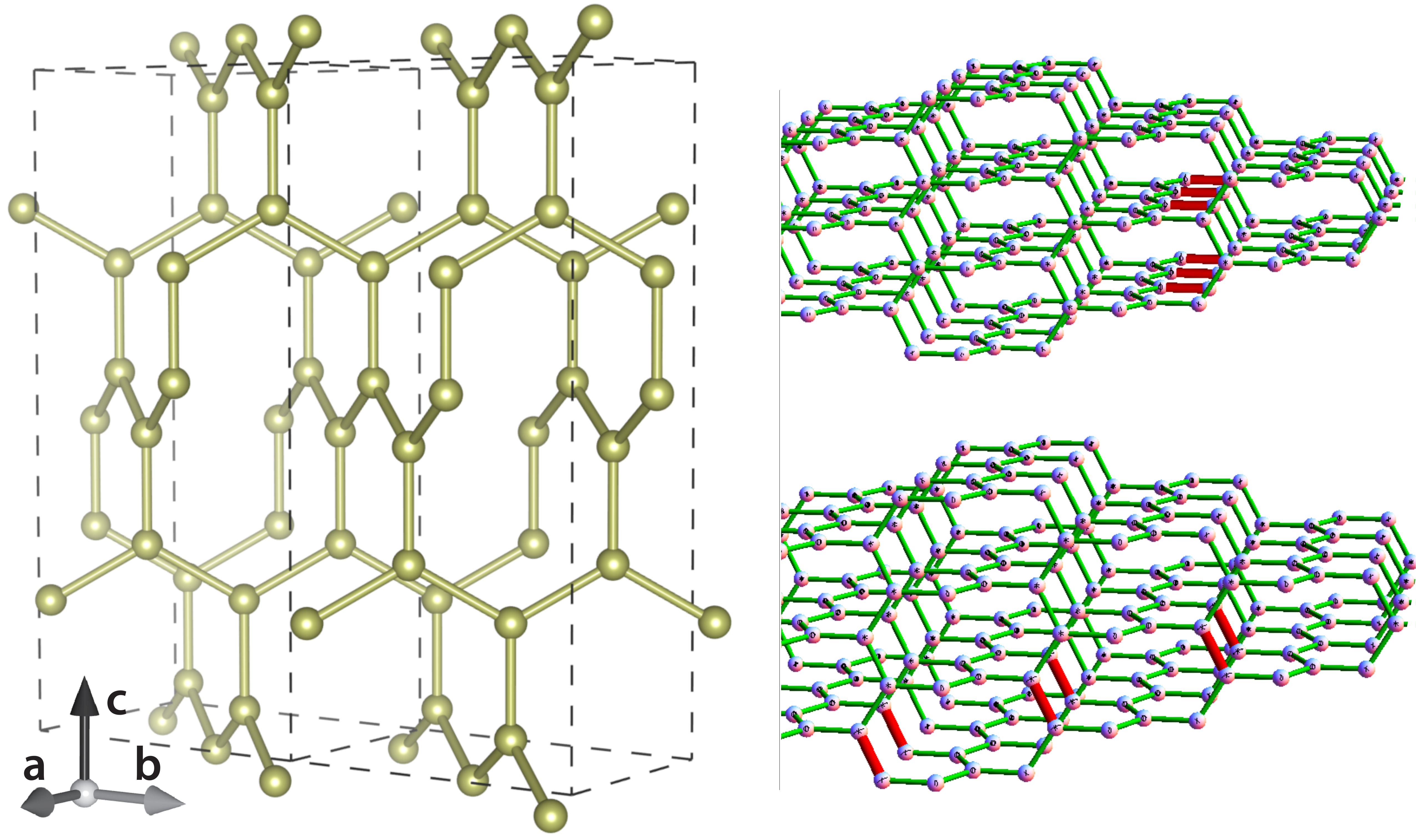}
\caption[]{{\bf The hyperhoneycomb lattice of Ir in $\beta$-\li213.} The hyperhoneycomb lattice (space group is \#70 \textit{Fddd}) has threefold coordinated sites and is the $n{=}0$ member of the harmonic honeycomb structural series. Its shortest loops are 10-site decagons, motivating the large-$\ell$ loop length approximation for solving the frustrated quantum Hamiltonian on the 3D lattice. Right: $\mathbb{Z}_2$ flux loops in the QSL phase. Selected bonds (dark orange) of type $z$ (top right) or $x,y$ (bottom right) are chosen to host nonzero vector potential $u_{i,j} = -1$ within the QSL $\mathbb{Z}_2$ gauge sector, producing a $\mathbb{Z}_2$ closed flux loop excitation, which encircles these bonds.}
\label{fig:hhc0}
\end{figure}

Recently, Li$_2$IrO$_3$ has been successfully synthesized in two insulating polymorph crystal structures consisting of  edge-sharing IrO$_6$ octahedra. 
In the $\beta$-Li$_2$IrO$_3$ polymorph, synthesized in powder form\cite{Takagi2014}, iridium ions form the 3D \textit{hyperhoneycomb lattice} as shown in Fig.~\ref{fig:hhc0}, with space group $Fddd$ (\#70). 
In the $\gamma$-Li$_2$IrO$_3$ polymorph, synthesized as single crystals\cite{analytis2014}, iridium ions form the \textit{stripyhoneycomb lattice} as shown in Fig.~\ref{fig:hhc1}, with space group $Cccm$ (\#66). 
Each of these three dimensional lattices is locally honeycomb-like, preserving threefold connectivity for every site.  Their unified geometry suggests an extension to a structural series, the ``harmonic honeycomb''  series\cite{analytis2014};  each structure in the series is labeled by an integer $n$, denoting the number of adjacent hexagon strips found in the lattice.   
In this notation, the \shc lattice $\gamma$-\li213 polymorph is the $n$=1 harmonic honeycomb iridate;  the hyperhoneycomb lattice  $\beta$-\li213  is the $n$=0 member; and the layered honeycomb $\alpha$-\li213 is described by $n$=$\infty$ (Tab.~\ref{tab:lattices}).

The $\gamma$-Li$_2$IrO$_3$ single crystals undergo a magnetic transition at about 38K, as evidenced by large anisotropic peaks in magnetic susceptibility\cite{analytis2014}. 
As also pointed out in the experimental analysis\cite{analytis2014}, the bond-aisotropic Kitaev-Heisenberg model Eq.~\ref{eq:KH1} is sufficient for capturing the large susceptibility anisotropy observed in experiment; within this scenario, large ferromagnetic Kitaev exchanges are necessary to fit the experimental data.  The susceptibility fit is shown in Fig.~\ref{fig:fits}; we elaborate on the magnetic couplings required for this fit in section \ref{sec:fits} below. We study the Hamiltonian Eq.~\ref{eq:KH1} with the parameters of the fit, classically as well as using the fully quantum large-$\ell$ approximation discussed below, and find in both cases a ground state with Stripy-X/Y magnetic order, shown in Fig.~\ref{fig:stripyX}.  In this pair of symmetry-related degenerate ground states, spins exhibit ferromagnetic correlations of spin component $S^x$ across Kitaev $x$-type  bonds or $S^y$ across $y$-type bonds.
Based on the susceptibility anisotropy we predict that these stripy magnetic  correlations occur in the low temperature phase of $\gamma$-Li$_2$IrO$_3$.
Indeed, since this work was presented, recently the magnetic order of $\gamma$-Li$_2$IrO$_3$ has been determined\cite{RaduSpiral} to be a counter-rotating noncoplanar spiral order in which the dominant spin correlations are exactly these  Stripy-X and Stripy-Y correlations, again requiring a magnetic Hamiltonian with strong FM Kitaev exchange.

In parallel with this work, a few other studies of 3D Kitaev-Heisenberg models have appeared.  Various properties of the hyperhoneycomb lattice model's magnetic phases and exact spin liquids were studied\cite{Kim2013b} while the magnetic phases at finite fields and temperature were explored using classical and semi-classical techniques \cite{Kim2013c}. The spin liquid was also studied at finite temperature using an Ising mapping of its Toric Code limit\cite{Motome2013}.  Another lattice related to the hyperkagome but with higher symmetry, dubbed the ``hyperoctagon lattice'', was introduced and the Kitaev spin liquid it supports was characterized \cite{Trebst2014}. 

Results here are complimentary to these studies, and are distinguished in three ways.  
First, we pull together the existing experimental results to make the case, based on single-crystal measurements, for strong Kitaev exchange in the 3D-\li213 materials.  
Second, we focus our attention on the hitherto-unstudied \shc lattice recently obtained as the structure of $\gamma$-\li213. Our magnetic Hamiltonians are informed by the experimental measurements and incorporate bond anisotropies dictated by the symmetries of the crystals.  Most  others\cite{Surendran2009,Kim2013b,Kim2013c,Motome2013} exclusively studied the hyperhoneycomb lattice, which we also study below. 
Third, in addition to studying the exactly solvable 3D spin liquids, we employ tensor product states -- higher dimensional generalizations of 1D matrix product states -- 
to obtain the fully quantum phase diagram in a large-$\ell$ limit.  The phase diagram we compute, for the frustrated quantum Hamiltonians motivated by the experiments, contains both magnetic and quantum spin liquid phases.  To our knowledge this is the first identification of a quantum spin liquid phase in a tree tensor network.

 \setlength{\tabcolsep}{6.5pt}
\begin{table}[]
  \centering
\begin{tabular}{|c|c|c|c|}
  \hline
  \multirow{2}{*}{Material} &   Harmonic- & \multirow{2}{*}{Lattice name} &  \multirow{2}{*}{Dim.}  \\
  & honeycomb \# &  &
 \\ \hline \hline
  $\alpha$-\li213 & $n=\infty$ & Honeycomb  & 2D \\ \hline
  $\beta$-\li213 & $n=0$  & Hyperhoneycomb & 3D \\ \hline
  $\gamma$-\li213 & $n=1$  & Stripyhoneycomb & 3D \\ \hline
\end{tabular}
  \caption{Iridates of the harmonic-honeycomb series: Ir lattice conventional name and dimensionality. We focus on  $\gamma$-\li213 single crystal measurements to extract magnetic Hamiltonians, which we study on the $\beta$- and $\gamma$- \li213 structures as well as on their tree tensor network approximation.}
  \label{tab:lattices}
\end{table}

\section{Summary of results} 
We begin (Sec.~\ref{sec:material}) by analyzing the relevance of the Kitaev interactions to \li213 using $\gamma$-\li213 single-crystal measurements. 
We discuss the interplay of chemistry and geometry in the A$_2$IrO$_3$ structures, aiming to understand the newly synthesized \shc and hyperhoneycomb lattices within a framework encompassing other 3D honeycomb lattices of edge-sharing IrO$_6$ octahedra.  We analyze in detail the argument, based on fitting magnetic susceptibility, that the magnetic properties of 3D-\li213 are captured by the bond-anisotropic Kitaev-Heisenberg model Eq.~\ref{eq:KH1}. Its key is the geometrical contrast between the  crystalline anisotropy, distinguishing the spatial $c$-axis,  and the magnetic susceptibility anisotropy, distinguishing the $b{=}z$ spin axis: the two are coupled by the $S^z S^z$ Kitaev exchange on $c$-bonds. We demonstrate this mechanism by analytically fitting the measured susceptibility to mean field theory of Eq.~\ref{eq:KH1}, and find (Fig.~\ref{fig:fits}) that it requires strong FM Kitaev exchange.

With this motivation for Eq.~\ref{eq:KH1} as a minimal Hamiltonian with dominant Kitaev exchange, we proceed (Sec.~\ref{sec:QSL}) to study its spin liquid phase in the Kitaev limit through the Majorana fermion exact solution. We extend the previous analysis of the hyperhoneycomb-graph Kitaev model\cite{Surendran2009}, and also analyze in detail the model on the \shc lattice of $\gamma$-Li$_2$IrO$_3$.  We compute the spin correlators as well as the spectrum of the emergent Majorana fermions, and find that the low energy excitations occur on a ring-like nodal contour, identical for the two 3D lattices. Introducing bond-strength anisotropy shrinks the nodal contour, and we find that the phase boundary between the gapped and gapless spin liquids is identical on all the finite-D lattices and independent of whether the bond-anisotropy breaks or preserves the lattice symmetries.

We give a simple but general counting argument based on the Euler characteristic formula that explicitly illustrates the lack of monopoles in (3+1)D $\mathbb{Z}_2$ lattice gauge theories, showing that closed flux loops rather than individual fluxes are the gauge-invariant objects.   The energy of the flux loop excitations is described not as a flux gap but rather by a loop tension, which we compute within the zero temperature exact solution to be $\tau = 0.011 |K|$ on both lattices. This tension combines with the extended nature of the loops to control the finite temperature behavior of the models, producing the finite temperature loop proliferation transition which confines the Majorana fermions. Together with the robustness of fermionic statistics (since flux attachment is impossible), this stability to finite temperature hallmarks the features unique to three dimensional fractionalization.
 
 Computing the quantum phase diagram of the full frustrated Hamiltonian is exponentially difficult; while such problems have been tackled in two dimensions, an unbiased phase diagram computation of the three dimensional model is currently impossible. We are able to capture it (Fig.~\ref{fig:radial_phases} and Sec.~\ref{sec:itebd}) by employing a limit inspired by the \hh  lattice, whose shortest loops are $\ell{=}10$ decagons.
Treating $\ell$ as a large control parameter and taking it to infinity, we reach the loopless Bethe tree lattice, which is infinite dimensional but preserves the key $z{=}3$ connectivity. This  $\ell{\rightarrow}\infty$ approximation is not analytically tractable, but rather admits an entanglement-based numerical solution using tensor product states (TPS). 
Gapped states can be efficiently represented as a TPS on a tree lattice (tree tensor networks); on the tree, as in 1D systems, the full entanglement between two halves of the system is carried by the single bond connecting them. We employ a TPS time evolving block decimation algorithm which works directly in the thermodynamic limit (iTEBD)\cite{Vidal2007}, which has been previously extended to the Bethe lattice for magnetic phases\cite{Nagaj2008,Nagy2012,Xiang2012} and other non-fractionalized phases\cite{Depenbrock2013,Su2013}. 
The iTEBD straightforwardly captures the FM and Neel magnetic orders as well as their duals\cite{Khaliullin2010,KHbhc}, the stripy and zigzag magnetic orders. 
 
However, quantum spin liquids are generally difficult to identify positively since they lack an order parameter.  Positive signatures can be elusive. Studies in 2D have relied on the sub-leading entanglement term known as the topological entanglement entropy\cite{Balents2012,Vishwanath2012a}, but this quantity is not defined nor computable on the tree lattice. Instead, we complement the TPS computation by analytically studying the gapped Kitaev QSLs on the loopless tree using the Majorana solution, computing the entanglement entropy from the fermion and gauge sectors on each bond as a function of anisotropy. 
We find that the TPS algorithm partially quenches the $\mathbb{Z}_2$ gauge field entanglement, utilizing the finite entanglement cutoff of the TPS representation to produce a minimally entangled ground state, and thereby circumventing the usual artifacts of the Bethe lattice. The resulting entanglement serves as a fingerprint which, alongside the vanishing magnetic order parameters, we use to identify the QSL phase within the iTEBD computation. 
Ours is the first positive-signature identification of a fractionalized quantum phase in the large-$\ell$ limit. 

This solution of the QSLs with their adjacent phases in the quantum large-$\ell$ approximation augments the ground state and finite temperature analysis within the solvable   three dimensional QSLs, yielding a remarkably complete picture of a fractionalized phase in a potentially realizable solid state system.


\section{Relevance of Kitaev interactions to the 3D lithium iridates}
\label{sec:material}
\subsection{Chemical bonding with IrO$_6$ octahedra} 
Oxides with octahedrally coordinated transition metals can bond in a variety of ways, sharing octahedral corners, edges, faces or a combination of these. 
Each bonding geometry results in a set of structures with various shared properties. 
Bond lengths are one such property, with nearest neighbor distances in iridates measuring ${\sim}3\AA$ in edge sharing compounds compared to ${\sim}4\AA$ in corner sharing ones. Symmetries are also correlated with bonding geometry; in corner sharing iridates where one oxygen is shared by exactly two iridia, four-fold symmetric structures generally arise, as in perovskite and layered perovskite structures.  Compounds with edge sharing octahedra also occur in a variety of related structures; the triangular lattice NaCoO$_2$, the hyperkagome Na$_4$Ir$_3$O$_8$ and the layered honeycomb Na$_2$IrO$_3$ are all examples. 

In the edge sharing iridates we consider, two oxygens are shared by exactly two iridia, and every iridium is coordinated by three others, belonging to a single plane. In a fixed coordinate system, there are multiple choices for the orientation of the triangular Ir-Ir-Ir plaquette, which are locally indistinguishable from the perspective of any given iridium atom. In general the octahedral symmetry will not be perfect and the distortion may favor the situation of the layered honeycomb compound Na$_2$IrO$_3$, where all of the Ir-Ir bonds lie within a common plane. 

However, for sufficiently high local symmetry approaching the full $O_h$ group, alternatives to the layered geometry become increasingly favorable.
Consider now the compounds with chemical formula Li$_2$IrO$_3$: this substitution of Na by Li is known to lead to much smaller distortions, since the Ir and Li ions are more similar in size. With the decreased octahedral distortion, multiple spatial orientations of the bonds should be more likely to occur.  This can result in complex structures, such as the \shc lattice of $\gamma$-\li213 and the hyperhoneycomb lattice of $\beta$-\li213. 

\begin{figure}[t]
\includegraphics[width=220 pt]{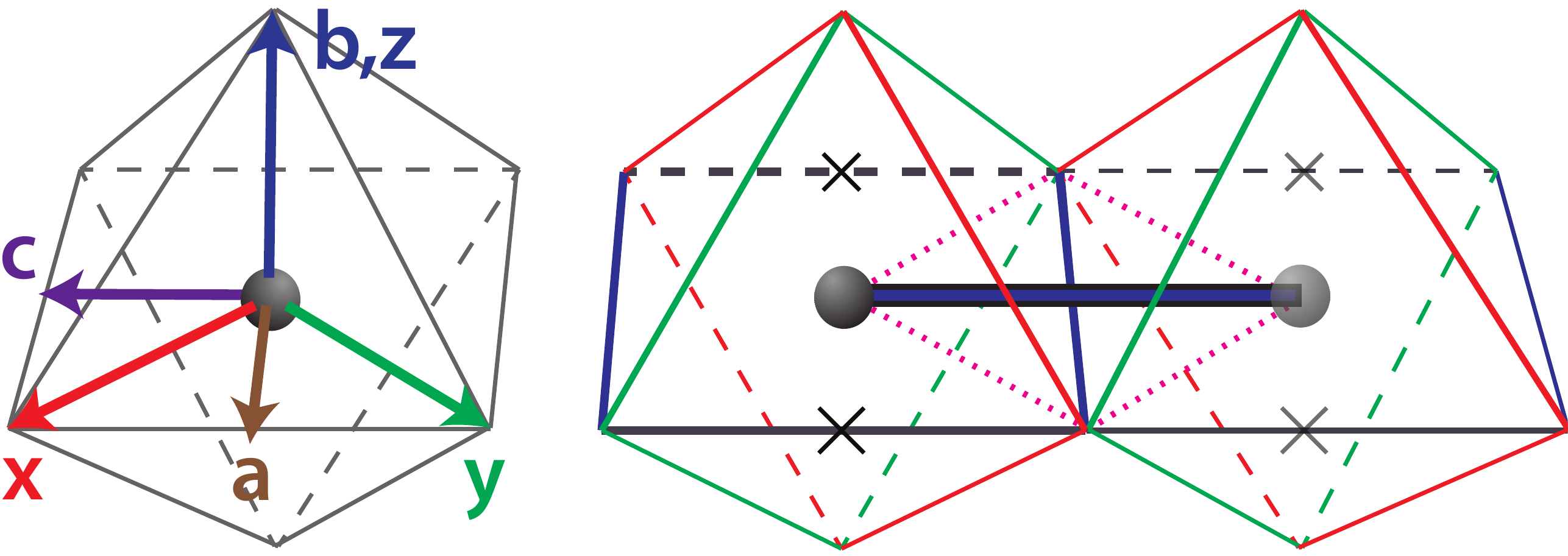}
\caption[]{{\bf Edge sharing IrO$_6$ octahedra in the 3D lattices. }  Iridium (purple sphere) is coordinated by six oxygens forming vertices of an octahedra. When octahedra share edges as shown, the exchange pathways (dotted purple lines) give rise to Kitaev interactions, coupling a spin component $\gamma\in\{x,y,z\}$  normal to that Ir-Ir bond and to the shared edge (shown in corresponding color $\{$red, green, blue$\}$). Octahedron at left shows the relation between the Ir-O $x,y,z$ axes and the crystallographic parent orthorhombic $a,b,c$ axes of the 3D lattices.  The symmetry-distinguished $c$-axis is also a preferred axis for Ir-Ir bonds (thick blue bond shown); the perpendicular edge (X'ed out gray lines) is not shared by any two IrO$_6$ octahedra. The $c$-bonds host $z{=}b$ Kitaev exchange.}
\label{fig:octahedra}
\end{figure}

\subsection{Symmetry and geometry of the harmonic honeycomb lattices}
Consider the harmonic honeycomb structures, which include all three currently known polymorphs of Li$_2$IrO$_3$. Except for the $n{=}\infty$ layered honeycomb with its vastly reduced crystal symmetry, these possess bonds with various orientations comprising \textit{all but one} of the possible orientations for edge-sharing octahedra.  This scenario is shown in  Fig.~\ref{fig:octahedra}: two opposite octahedra edges are forbidden from bonding, and distinguish the spatial direction $c$, parallel to these edges. The other edges on the same square create Ir-Ir bonds lying along the $c$ axis, resulting in the bond anisotropy described above.  The $c$ axis is thus distinguished for all harmonic honeycomb lattices; this is also reflected in the symmetry properties of each particular lattice.  For example in the \shc lattice, the space group \textit{Cccm} has a single mirror plane, whose normal is the $c$ direction.  This unifying feature also suggests that a single global orthorhombic $a,b,c$ parent coordinate system can describe the various lattices, as is indeed true. 
The vectors of these parent orthorhombic axes, as well as explicit coordinates for the \shc and \hh lattices, are given in Appendix~\ref{sec:vectors}. 

Recall that Kitaev spin coupling along the Bloch sphere Cartesian axis $\gamma\in\{x,y,z\}$ occurs\cite{Khaliullin2009} for the four octahedra edges (and associated Ir-Ir bonds) whose plane is normal to the spatial Cartesian axis $\gamma$. 
The relation between the $a,b,c$ crystallographic axes of Fig.~\ref{fig:hhc1} and the octahedral Ir-O Cartesian axes, as shown in  Fig.~\ref{fig:octahedra}, is $
\{\hat{a},\hat{b},\hat{c}\}=\left\{(\hat{x}{+}\hat{y})/\sqrt{2}, \hat{z}, (\hat{x}{-}\hat{y})/\sqrt{2}\right\} $. 
   The $c$-bonds (i.e. the bonds lying along the $c$ axis) carry Kitaev coupling of spin component $\hat{b}=\hat{z}$, and will also be denoted interchangeably by their Kitaev label, with the notation ``$z$-bonds''. The remaining bonds on the lattice (``$d$-bonds''), which are all related to each other by symmetries, carry Kitaev labels $x$ and $y$. For the \shc lattice, the $c$-bonds are further distinguished into two types, those within hexagons and those between hexagons, which are themselves not related to each other by symmetry. For the sake of simplicity here we have not introduced additional parameters in the Hamiltonian to distinguish these two $c$ bond types, as we expect such bond strength anisotropy between the different $c$ bonds to be a secondary effect.

\subsection{Capturing $\gamma$-\li213 susceptibility with bond-anisotropic Kitaev interactions}
\label{sec:fits}
The symmetry distinction between $z$ and $x,y$ type bonds implies that if the Kitaev coupling is strong, the magnetic susceptibility should have a distinctive $z$-axis response compared to its $x,y$ axes response, at least at temperatures above the magnetic transition.  If the Kitaev coupling $K_c$ on the $z$-bonds is more ferromagnetic than the Kitaev coupling $K_d$ on the $x,y$-bonds, it suggests an anisotropic susceptibility with larger response along $z$.  Exactly such an anisotropy is observed in the $\gamma$-\li213 experiment\cite{analytis2014}. However, to preserve the strong $z$-axis susceptibility which is observed also below the ordering transition, the resulting magnetic order should not have any significant spin component aligned along the $z$ axis.  This places a condition on the magnetic coupling, to disfavor magnetic order alignment along $z$, which is partially at odds with the condition necessary to favor susceptibility anisotropy with large $\chi^z$. 

\begin{figure}
\includegraphics[width=\columnwidth]{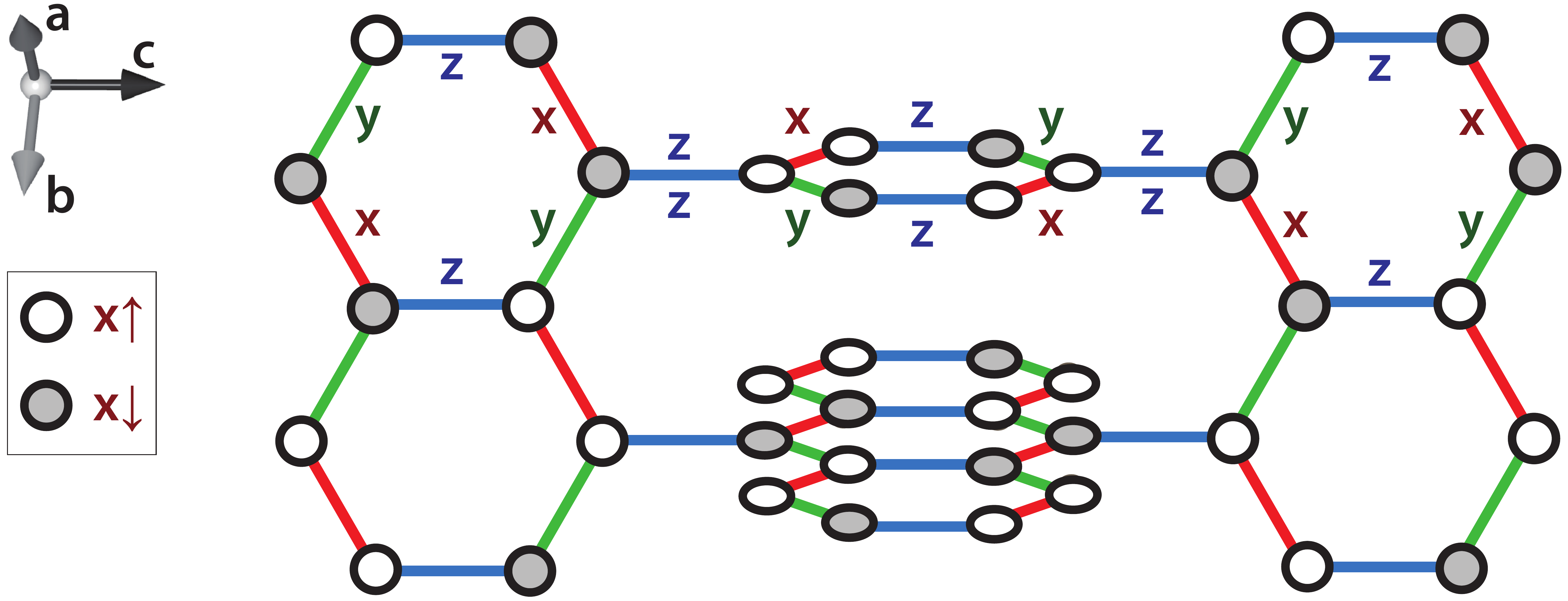}
\caption[]{{\bf Stripy-X magnetic pattern on the \shc lattice.} 
The classical magnetic pattern associated with the \textit{Stripy-X/Y} AF ordered phase, here shown for Stripy-X correlations. Spins, collinear along $S^x$ (white/gray sites correspond to $S^x$ up/down spins),  are aligned along $x$-type (red) bonds and anti-aligned along $y$-type (green) and $z$-type (blue) bonds. 
The unit cell is doubled (ordering wavevector $(\pi,\pi,0)$) to form the full unit cell of the parent orthorhombic $a,b,c$ axes.  These Stripy-X/Y correlations are predicted by the strong FM Kitaev exchanges necessary for the mean field fit to the $\gamma$-\li213 magnetic susceptibility.}
\label{fig:stripyX}
\end{figure}

To achieve strong anisotropy in the susceptibility $\chi$, the Heisenberg couplings must be small compared to the anisotropic single-spin-component exchanges, in this case the large ferromagnetic Kitaev exchanges. Since the low temperature phase is not a ferromagnet, the Heisenberg couplings should be antiferromagnetic.  This region of parameter space hosts two types of magnetic order, \textit{Stripy-Z} and \textit{Stripy-X/Y}, with different symmetry properties. With no additional anisotropies, \textit{Stripy-Z} nominally hosts spins aligned along the $z$ axis and can thus be ruled out. A more general property of the \textit{Stripy-Z}  phase is that, because of the two symmetry-inequivalent $z$-type bonds of the \shc lattice, it should generically exhibit a nonzero net moment. 
We therefore focus on Hamiltonians within the \textit{Stripy-X/Y} phase (Fig.~\ref{fig:stripyX}), as the simplest ``minimal order'' which is consistent with magnetic susceptibility and captured by the minimal Hamiltonian Eq.~\ref{eq:KH1}. We expect additional small exchange terms to modify the ground state order, but preserve the Stripy-X/Y correlations of this minimal phase.

The constraints on the couplings can be seen explicitly by treating the Hamiltonian classically, and extracting susceptibility by mean field theory (details are given in Appendix~\ref{sec:meanfield}). The magnetic interactions of  Eq.~\ref{eq:KH1} were supplemented by a g-factor tensor chosen to match the susceptibility at the highest temperatures measured, with principal values $g^{x+y}{=}g^z{=}1.95$, $g^{x-y}{=}2.35$. 
Within the mean field treatment of the  \textit{Stripy-X/Y} phase (in the regime $K<0$, $J>0$), the transition temperature is given by  $T_N=(J_c+|K_d|)/4$. The susceptibility peaks at this temperature, taking values 
\begin{align}
\chi^{b b}(T_N)&=(g^b)^2 \mu_B^2 / (2(J_c + J_d) - (|K_c| - |K_d|) )\\
\chi^{a a}(T_N)&=(g^a)^2 \mu_B^2 / (2(J_c + J_d)) \nonumber
\end{align}
and with $\chi^{c c}$ similar to $\chi^{a a}$. The observed susceptibility anisotropy then suggests a large value for the difference $|K_c| - |K_d|$.  However, the stability of the Stripy-X/Y phase against Stripy-Z order is controlled by the constraint 
\begin{equation}
|K_c| - |K_d| < 2(J_c - J_d) .
\end{equation}
There is a finite window of parameters which fit the data within these analytical constraints.  One possibility for the couplings, as shown in Fig.~\ref{fig:fits}, is (in meV): $K_c=-17, K_d=-7, J_c=6.3, J_d=0.8$.  The Hamiltonian with this set of parameters was also studied beyond the classical limit, using tensor product states within the infinite dimensional large-$\ell$ approximation, and determined to lie within the \textit{Stripy-X/Y} quantum phase. 

This parameter regime of the fit, large ferromagnetic Kitaev exchange and small antiferromagnetic exchange, is consistent with Jackeli and Khaliullin's original proposal\cite{Khaliullin2009,Khaliullin2010} and with the recent \na213 ab initio\cite{Imada2014}.  The extent of the anisotropy is qualitatively similar to the \na213 ab initio prediction as well; the parameters computed for \na213 are\cite{Imada2014} $K_c=-30.7, K_d=-23.9, J_c=4.4, J_d=2.0$ meV, and larger anisotropy is expected for the \shc lattice because the special $c$ bonds directly form the special axis of its \textit{Cccm} space group. 

\subsection{Necessity of large Kitaev interactions for describing magnetic measurements on $\gamma$-\li213}
The analysis in the previous section showed analytically that within mean field theory, fitting the observed anisotropic susceptibility required a large ferromagnetic Kitaev exchange, dominant over a smaller AF Heisenberg exchange.  The bond-dependent Kitaev interactions $K_c,K_d$ then capture the observed susceptibility at temperatures both above and below the $\sim 40$ K magnetic transition.

The primary conclusion of this analysis is the argument that the bond-anisotropic Kitaev-Heisenberg Hamiltonian is appropriate for describing current experimental data on 3D-\li213 and requires quite large Kitaev exchanges $|K|>>J$.  The nominal ground state of the fitted Hamiltonian is the simple collinear phase Stripy-X/Y, but in the real crystal we do not expect the spin direction to be locked to $\hat{x}$ or $\hat{y}$, but rather  expect it to sample across the $(a,c)$ (or equivalently $(x,y)$) plane. A secondary conclusion is therefore a susceptibility-based prediction for the low temperature magnetic pattern of $\gamma$-\li213, namely presence of the \textit{Stripy-X/Y} correlations of the fitted Hamiltonian. 

As mentioned in the introduction, this 4-parameter fit to magnetic susceptibility, with parameters $(K_c,K_d,J_c,J_d)=(-17,-7,6.3,0.8)$ meV, is consistent with the 6-parameter Hamiltonian which captures the noncoplanar spiral magnetic order which has just been recently observed\cite{RaduSpiral} in $\gamma$-\li213. That 6-parameter fit supplements Eq.~\ref{eq:KH1} by $c$-axis Ising exchange $I^c_c$ on $c$-bonds and $J_2$ Heisenberg exchange on second-neighbors, and gives the values $I^c_c{=}{-}4.5,J_2{=}{-}0.9$, $(K_c,K_d,J_c,J_d)=(-15,-12,5,2.5)$ meV. The associated Stripy-X and Stripy-Y correlations expected for such a quantitatively similar Hamiltonian are also observed in the complex spiral order.  
Most importantly, we observe that in each analysis, independently, large and FM Kitaev exchanges $K_c,K_d$  are necessary to describe the material.

\section{Quantum spin liquids in three dimensions}
\label{sec:QSL}
Let us now tune the Heisenberg couplings $J$ to zero, taking the limit of a pure Kitaev Hamiltonian. Though this limit does not describe the experiments on \li213, it offers a wide range of interesting phenomena associated with 3D fractionalization, which may turn out to be experimentally accessible at a future date. 

\subsection{Solution via Majorana fermion mapping}
Kitaev's solution\cite{Kitaev2006} of the honeycomb spin model relies on a local condition --- each site touches three bonds carrying the three different Kitaev labels --- and hence may be generalized to lattices with $z=3$ coordination number. 
In order to discuss important subtleties which will arise later (in infinite dimensions), let us briefly review the solution here. 
 The $S=1/2$ algebra is represented in an enlarged Hilbert space via four majorana fermions
\begin{equation}
S_i^a \rightarrow \frac{1}{2} i \chi_i^0 \chi_i^a , \ \ \{\chi_i^a,\chi_{i'}^{a'}\} = 2\delta_{i,i'}\delta_{a,a'}.
\end{equation} 
The enlarged Hilbert space Kitaev Hamiltonian $\tilde{H}$ is then a free majorana fermion $\chi^0$  minimally coupled 
to a $\mathbb{Z}_2$ vector potential $a_{i,j}$ with $e^{i\pi a_{i,j}}\equiv u_{i,j}=i \chi_i^{\gamma_{i j}} \chi_j^{\gamma_{i j}}$ living on links $\langle i j \rangle$. The gauge field operators $u_{i,j}$ all commute with each other and with $\tilde{H}$, so $\tilde{H}$ may be diagonalized by solving a free fermion problem for each gauge field configuration $\{u_{i,j}=\pm 1\}$. 
Here $u_{i,j}$ is identified as a gauge field because, while in the enlarged Hilbert space it is a simple $\mathbb{Z}_2$ bond variable set by the majorana fermion occupancy, there is a set of site operators $D_i \equiv \chi_i^0 \chi_i^1 \chi_i^2 \chi_i^3$ which are the identity within the physical spin Hilbert space but act as a lattice gauge transformation on the link variable $u_{i,j}$. 
Projection to the physical spin Hilbert space is implemented by symmetrizing over all local gauge transformations $D_i$, with the projection operator $P=\prod_i (1+D_i)/2$. 

\begin{figure}[t]
\includegraphics[width=140 pt]{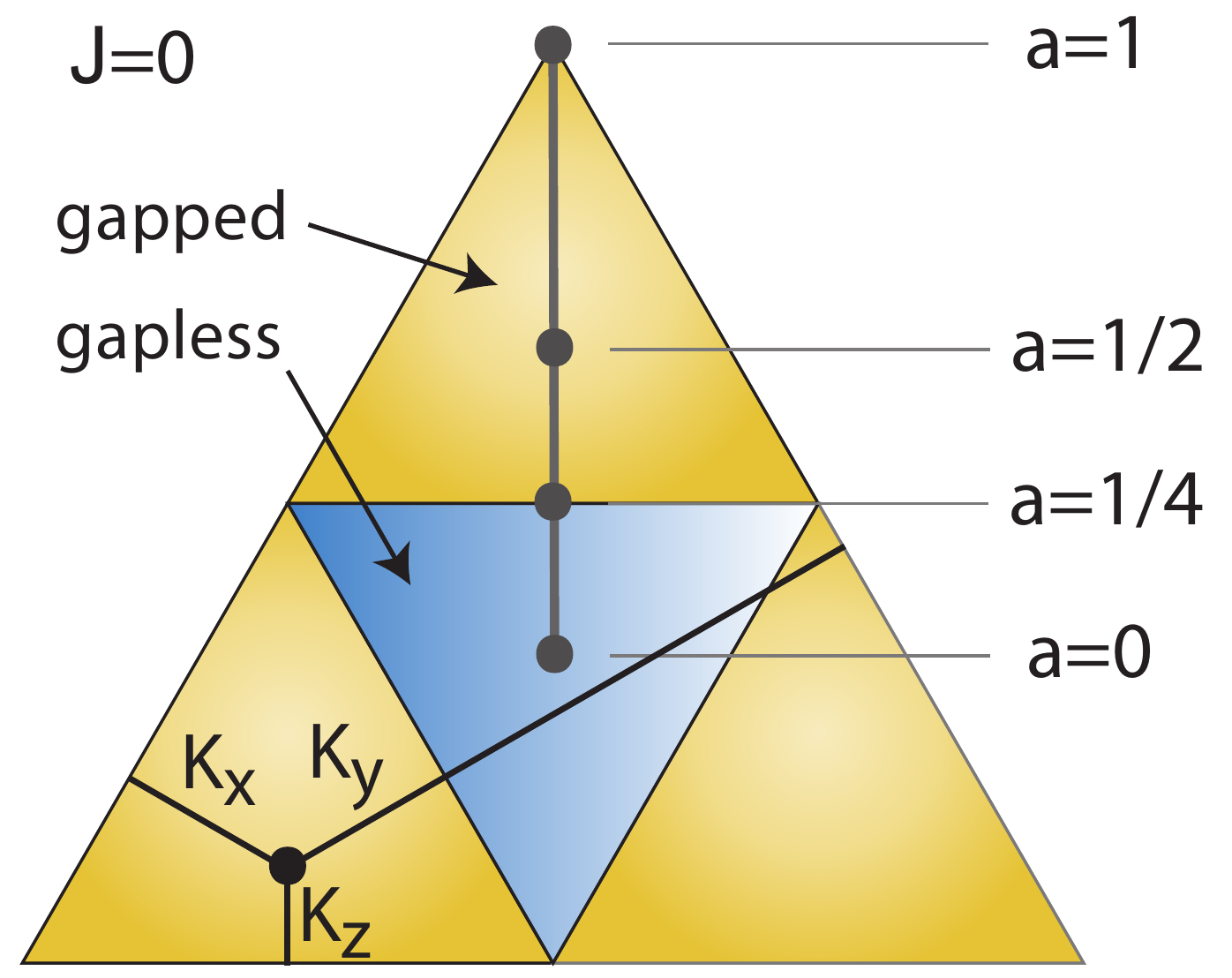}
\caption[]{{\bf Kitaev spin liquid with bond anisotropy. } 
The phase diagram of the 3D Kitaev spin liquids as a function of bond anisotropy, designating whether the emergent Majorana fermions are gapped or gapless, is 
identical on the 3D lattices and the 2D honeycomb model, and is independent of whether the anisotropy breaks or preserves lattice symmetry. 
Here the magnitude of Kitaev coupling $|K_x|,|K_y|,|K_z|$ is given by the distance to the respective edge of the triangle. The vertical line corresponds to the symmetry-allowed bond-anisotropy, modifying $|K_z|$ by the parameter $a$ (Eq.~\ref{eq:KH2}) with particular values shown. }
\label{fig:triangle}
\end{figure}
 
Let us now discuss the consequences of this majorana fermion solution for the 3D trivalent lattices. Some of the phenomenology was previously explored\cite{Surendran2009} for a 3D lattice whose connectivity graph matches the hyperhoneycomb's. 
 The projection from gauge to physical Hilbert space is aided by gauge invariant operators whose eigenvalues, commuting with the Hamiltonian, label physical sectors of states. These closed Wilson loops are the usual $\mathbb{Z}_2$ fluxes piercing the elementary plaquettes. Flipping $u_{i,j}$ on a bond inserts flux in adjoining plaquettes. 
As discussed earlier, the 3D lattices possess symmetries as well as graph connectivity which distinguish one bond type, $z$, from the other two bond types $x$ and $y$. On the hyperhoneycomb lattice, flipping $u_{i,j}$ on a $z$ type bond creates fluxes on the eight adjacent plaquettes, while flipping $u_{i,j}$ on $x$ or $y$ type bonds changes the flux on the only six adjacent plaquettes. On the \shc lattice,  the elementary plaquettes come in multiple forms, consisting of $\ell=6$ hexagons together with larger $\ell=14$ plaquettes.

\subsection{Extended flux loop excitations in the 3D QSL}
The gauge field sector on the 3D-lattices Kitaev models is a 3+1D $\mathbb{Z}_2$ lattice gauge theory\cite{Kogut1979}. The product of $u_{i,j}$ around a minimal closed contour gives the flux through an elementary plaquette of the lattice. The product of fluxes on plaquettes surrounding an elementary volume element multiplies to the identity: this is equivalent to the fact that there are no magnetic monopoles in the $\mathbb{Z}_2$ theory. Each elementary volume carries a zero monopole charge and thus acts as a constraint, forcing the number of flux lines piercing the volume to be even. These constraints ensure that the flux lines only appear within closed flux loops.  

It is important to note that while in the 2D honeycomb case the magnetic fluxes are the gauge invariant result of projecting the gauge theory, in the case of three spatial dimensions, individual fluxes are not gauge invariant. Rather, only closed flux loop configurations are the physical gauge invariant excitations of the model. The individual fluxes cannot be gauge invariant labels of sectors of the Hamiltonian since they don't even correctly count the physical degrees of freedom of the gauge theory.  The constraint of closed loops fixes this counting; the closed magnetic loop configurations exactly label the gauge invariant sectors of the $\{u_{i,j}\}$ after projection. 

This can be seen as follows (explicitly verifying this statement in the \shc and \hh lattices is also straightforward).  Consider the lattice with periodic boundary conditions (rigorously it is a CW-complex topologically equivalent to the 3-torus), and count the number of cells of every dimension --  sites, bonds, plaquettes and enclosed volumes. The Euler characteristic formula (as generalized by homology theory) then shows that
\begin{equation} 
N_\text{sites} - N_\text{bonds} + N_\text{plaquettes} - N_\text{volumes} = 0 .
\end{equation}
 The combination $N_\text{plaquettes} - N_\text{volumes}$ is of interest here, since every plaquette is associated with a flux, but each enclosed volume presents a condition on the adjacent fluxes (they must multiply to the identity). This constraint, due to the lack of monopoles in the $\mathbb{Z}_2$ gauge theory, is responsible for the flux lines forming closed flux loops. The number of independent such loops is given by the number of possible flux lines minus the number of constraints, ie $N_\text{flux loops} = N_\text{plaquettes} - N_\text{volumes}$.  Furthermore let us restrict to our case of interest where sites have coordination number $z=3$ and so $N_\text{bonds} = (3/2)N_\text{sites}$. Then the formula becomes 
\begin{equation}
N_\text{flux loops} = N_\text{sites}/2
\end{equation}
as required: the gauge field flux sector hosts half of the spin degrees of freedom, while the majorana fermion particle sector hosts the other half. 

This observation implies the following important fact:  while in the 2D Kitaev model, the flux sector is described by a gap to flux excitation, this is not the correct description for this 3D model.  Rather, in 3D, the fluxes form closed loops, of arbitrary size.  These loops possess a loop tension. The gap for a loop of a particular length is found by multiplying its length by the loop tension.  We have computed a numerical value for this loop tension, as further discussed below.

 Lieb's flux phase theorem\cite{Lieb1994}, which shows that the 2D honeycomb ground state has zero flux per $\ell=6$  hexagon,  suggests that the $\ell=6$, $\ell=10$ and $\ell=14$  loops of the  \shc and \hh lattices, whose length is equal to $2$ mod $4$, 
 should also carry zero flux in the ground state. We have checked numerically that the ground state on small finite systems lies in the sector with no flux loops.

\subsection{Majorana fermion excitations}
The Kitaev QSL possesses emergent quasiparticles which are fermionic, arising out of the interacting bosonic spin model. The emergent fermions are as real as physical electrons, but carry no usual electric charge; and moreover are Majorana (self-adjoint), related to the particle-hole-symmetric excitations of superconductors. As in the 2D honeycomb model, in which the fermion dispersion possesses graphene's Dirac nodes, the fermionic dispersion in the 3D lattices is gapless for the isotropic model. The sublattice symmetry present in all the 3D harmonic honeycomb lattices ensures that time reversal remains a symmetry in the QSL phase, and the Majorana fermion spectrum is particle-hole symmetric. Similarly to the graphene-like Dirac cones appearing in the Majorana spectrum of the 2D Kitaev honeycomb model, where the 0D point-like nodes carry codimension of 2, the 3D Kitaev models can host gapless excitations along 1D nodal lines within the 3D Brillouin zone.

\begin{figure}[t]
\includegraphics[width=150 pt]{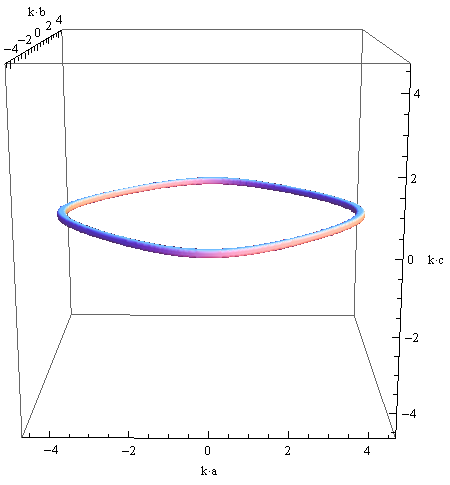}
\caption[]{{\bf Nodal contour of Majorana fermions in the 3D QSL. } In the gapless 3D Kitaev spin liquid phase, the emergent Majorana fermions are gapless at momentum points which form this 1D contour within the 3D momentum space.
The contour is identical for the QSLs on the \shc lattice and on the \hh lattice; it is set by $\vec{k}\cdot \vec{c} = 0$  and $\cos \left( \vec{k} \cdot
\vec{a}/ 2 \right)+\cos\left( \vec{k} \cdot \vec{b}/ 2 \right) =
1/2$.  with $a,b,c$ the parent orthorhombic axes (shown). 
  Introducing coupling-strength bond anisotropy shrinks this contour until it collapses to a point and then gaps out, yielding the gapped spin liquid phase. 
} 
\label{fig:nodes}
\end{figure}

The spectrum of Majorana fermions is computed in Appendix \ref{sec:Mspectrum}, and turns out to be identical on both 3D lattices. It is formed by momenta $k$ satisfying the two equations $\vec{k}\cdot \vec{c} = 0$  and $\cos \left( \vec{k} \cdot
\vec{a}/ 2 \right)+\cos\left( \vec{k} \cdot \vec{b}/ 2 \right) =
1/2$.  This set of momenta form a closed 1D ring-like contour of gapless excitations, lying within the BZ interior, which is plotted in Fig.~\ref{fig:nodes}.
Indeed this is the dispersion of a nodal 3D superconductor: the Majorana fermions are gapless along a 1D ring of points in the 3D momentum space, forming  a superconductor line node which here happens to close into a ring within the interior of the first Brillouin zone. 

Within each sector with its associated flux loop configuration, we may study how the Majorana fermions propagate. 
The fermions are charged under the gauge field, and hence interact with the magnetic loop excitations through an Aharonov Bohm effect, analogous to that occurring between electrically charged electrons and conventional E\&M magnetic flux lines. The interaction is as follows: when a fermion winds through the interior of a magnetic flux loop, it encircles one flux line and receives a minus one $(-1)$ phase to its single particle wave function. 

\subsection{Spin-spin correlators}
The spin-spin correlators at equal time may be computed straightforwardly within the fermion mapping; as in 2D, they are\cite{Kitaev2006} only nonzero between spins on nearest-neighbor sites and then only between spin components matching that bond's Kitaev label. Hence the nonzero spin correlators $G$ are also equivalently the energy $E$ carried by the bond (divided by the coupling), specifically $G=E/K$.  For notational simplicity we quote correlators $G$ for $K<0$, in which case the correlators are positive; for $K>0$, correlators simply gain a minus sign. Here we report results at the isotropic point of the Hamiltonian, though of course lattice symmetry still comes into play. We find that the average bond correlator (again, proportional to the energy per bond) is $G^{3D;0}_0 = -0.1284 $ for the 3D \hh and $G^{3D;1}_0 = -0.1290 $ for the 3D \shc, only 2\% higher than the 2D honeycomb result\cite{Kitaev2006} $G^{2D}_0 = -0.1312 $. 

For the \hh lattice, the $z$ bonds and $x,y$ bonds correlators are
\begin{equation} G_z= 0.1314, \quad G_{x,y}=0.1268 . \end{equation}
The \shc lattice has two symmetry-distinct types of $z$ bonds: those within hexagons (``$z[h]$'') and those within length-14 loops (``$z[f]$'') . 
The correlators are
\begin{equation} G_{z[h]}=0.1337 , \quad G_{z[f]}=0.1269, \quad G_{x,y}=0.1283 . \end{equation}
The large correlations on hexagon-$z$ bonds could be explained as strong resonances within a hexagon, combined with a lattice symmetry effect that, for both the \hh and the \shc lattices, gives stronger correlations on the special-axis $z$ bonds. Surprisingly, this global symmetry effect is almost as powerful as the hexagon resonances: it produces $x$,$y$-bond correlators which are only slightly stronger than those on the cross-hexagon-stripe $z[f]$ bonds. 

\subsection{Nodal contour under bond-strength anisotropies and broken symmetries}
Increasing the coupling strength $K$ on one bond type is an anisotropy which preserves exact solvability of the model, in 3D as well as 2D. Increasing $K$ on bonds of one Kitaev-type  shrinks down the nodal contour, until it vanishes and gaps out the fermions when the Kitaev exchange for any one bond type becomes larger than the sum of the other two. However, consider that the \hh and \shc lattice symmetries already distinguish $z$-bonds and their axis $\hat{c}$ as a special direction; increasing the strength of $z$ bonds is an anisotropy which is generically expected to arise given the crystal symmetries. 

Increasing the strength of Kitaev exchange on $z$-bonds shrinks the nodal contour towards its center at the Gamma point $\vec{k}=0$. With sufficient anisotropy, the contour collapses at $\vec{k}=0$ and then disappears, producing a gapped Majorana fermion spectrum (Fig.~\ref{fig:triangle}).  But anisotropies for $x$ or $y$ type bonds do break a symmetry of the isotropic model. When increasing bond strength on $x$ or $y$ bonds, the nodal contour goes through a Van Hove singularity as it expands to touch the BZ surface, and then becomes centered around a BZ corner, towards which it gradually shrinks. This transition through a Van Hove singularity is an aspect associated with breaking crystalline symmetries.   However, while these aspects of the nodal contour are different between symmetry-breaking ($x,y$) and non-symmetry-breaking ($z$) anisotropy, the resulting phase diagram of the spin liquid phase is the symmetric diagram shown in Fig.~\ref{fig:triangle}, identical to that of the symmetric 2D honeycomb lattice. 

Each of the 3D lattices supports two different types of limits of large anisotropy, $z$ and $x/y$ types, which are associated with different three dimensional Toric Code models living on different $z=4$ lattices.
Each of these 3D Toric Code models is a pure $\mathbb{Z}_2$ gauge theory, with commuting plaquette terms formed by sites on a particular $z=4$ lattice set by the type ($z$ or $x,y$) of anisotropy. The Toric Code lattices are easily constructed by collapsing the strong-coupled bond into a site. 
The Toric Code  $\mathbb{Z}_2$ flux operators act on plaquettes of reduced size: the \hh decagons turn into Toric Code hexagon plaquettes, while \shc hexagons (as in 2D) turn into Toric Code square plaquettes. 

\subsection{Gap via breaking of time reversal}
Breaking time reversal symmetry with an external magnetic field induces oriented imaginary second neighbor hopping of the majorana fermions. The sign (orientation) of this imaginary hopping, necessary for majorana fermions, is set (as in 2D\cite{Kitaev2006}) by the sign of the permutation of the two Kitaev bond labels traversed. We find that breaking time reversal fully gaps out the entire majorana nodal contour. 

Interestingly, though, special behavior emerges at ultra-low fields. To lowest order, the external field introduces a mass gap which changes sign across the nodal contour, leaving two gapless band-touching points. At the next order of the external field, these points are gapped out as well; but they may control the physics at low fields and low energy scales.

\subsection{Fractionalization in 3D: extended loops and finite temperature confinement transition}
Enlarging spatial dimensionality from two to three dimensions changes the nature of the spin liquid phase; the two most interesting differences involve fermions and finite temperature. 
In the two dimensional spin liquid away from the exactly solvable point\cite{Kitaev2006}, the flux excitations gain dynamics and interact with the majorana fermions; the low energy excitations could then be bound fermion-flux pairs, composite particles with simple bosonic statistics. 
In contrast, consider the three dimensional spin liquid; 
here fluxes are not pointlike particles but rather closed magnetic loops, so the emergent fermions cannot merely bind a (point-like) flux to transmute into bosons, and thus their 3D fermionic statistics are more robust. 
While fermions can e.g. bind into Cooper pairs to disappear from the lowest energy theory, a fundamental excitation in the model still necessarily preserves fermionic statistics. The fermions remain until a phase transition either confines them or transmutes them into bosons via a more complicated mechanism such as that recently explored in transitions between symmetry protected topological phases\cite{Wang2013}.

Three dimensional spin liquid phases generally admit a key characteristic distinguishing them from 2D spin liquids: the 3D spin liquid phases survive to finite temperatures. Such is true for the Kitaev 3D spin liquid phase, which undergoes a distinct entropy-driven phase transition to a classical paramagnet. In 2D QSLs a finite density of fluxes exists at any nonzero temperature; the fermions gain a phase of $(-1)$ when encircling each of the fluxes and the resulting destructive interference results in a $T=0$ confinement transition to the paramagnet phase. But in the 3D QSL, magnetic fields appear in extended loop excitations, whose energy is proportional to their length via an effective loop tension. The loop energy diverges with its length. At finite temperature there is a finite density only of short loops, whose small cross-sectional area renders them invisible to the fermions. A finite probability for flux-encircling paths occurs only with macroscopically large loops, which cost diverging energy and hence appear at vanishing density. Entropy however favors longer loops, and so the free energy at finite temperature $T$ for a loop of length $L$ appears as (for long $L$)
\begin{equation}
F(L;T) = \left(\tau + \delta \tau (T) -\tilde{s} T \right) L
\end{equation}
where $\tilde{s}$ is the entropy contribution to the loop tension, roughly the natural logarithm of the coordination number of the dual lattice (where magnetic loops live), $\tilde{s} \approx \text{log}(z_\text{dual})$; $\tau$ is the zero temperature flux loop tension; and $\delta \tau (T)$ is the contribution to the effective loop tension at finite temperature due to interactions mediated by the gapless fermions. 

Because the entropy is likely the dominant contribution and appears with a negative sign, the effective magnetic loop tension renormalizes to lower values at finite temperature.  At a temperature $T_c$ the tension becomes negative and proliferates arbitrarily large magnetic loops in a transition analogous to Kosterlitz-Thoughless flux unbinding, which then confine the fermions.  We estimate the critical temperature $T_c$ by computing the zero temperature value of the magnetic loop tension $\tau$ in the isotropic Hamiltonian, finding the result 
\begin{equation} \tau = 0.011 |K| \end{equation}
for both \shc and \hh in different geometries and for different loops roughly independent of the loop shape, underlying bond/plaquette type, and for large loop lengths of up to 30 cross-sites (on the hyperhoneycomb lattice e.g. Fig.~\ref{fig:hhc0}), implying the estimate  $T_c\sim |K|/100$.

\section{Quantum phase diagram in an infinite-D approximation}
\label{sec:itebd}
The Kitaev-Heisenberg model suffers from the ``sign problem'' of frustrated quantum Hamiltonians: unbiased algorithms for computing its phase diagram require computational costs scaling exponentially with system size, a problem greatly exacerbated in a three dimensional lattice. Unbiased reliable computations of the phase diagram on the three dimensional lattices  are not possible at present time.

\begin{figure}[t]
\includegraphics[width=200 pt]{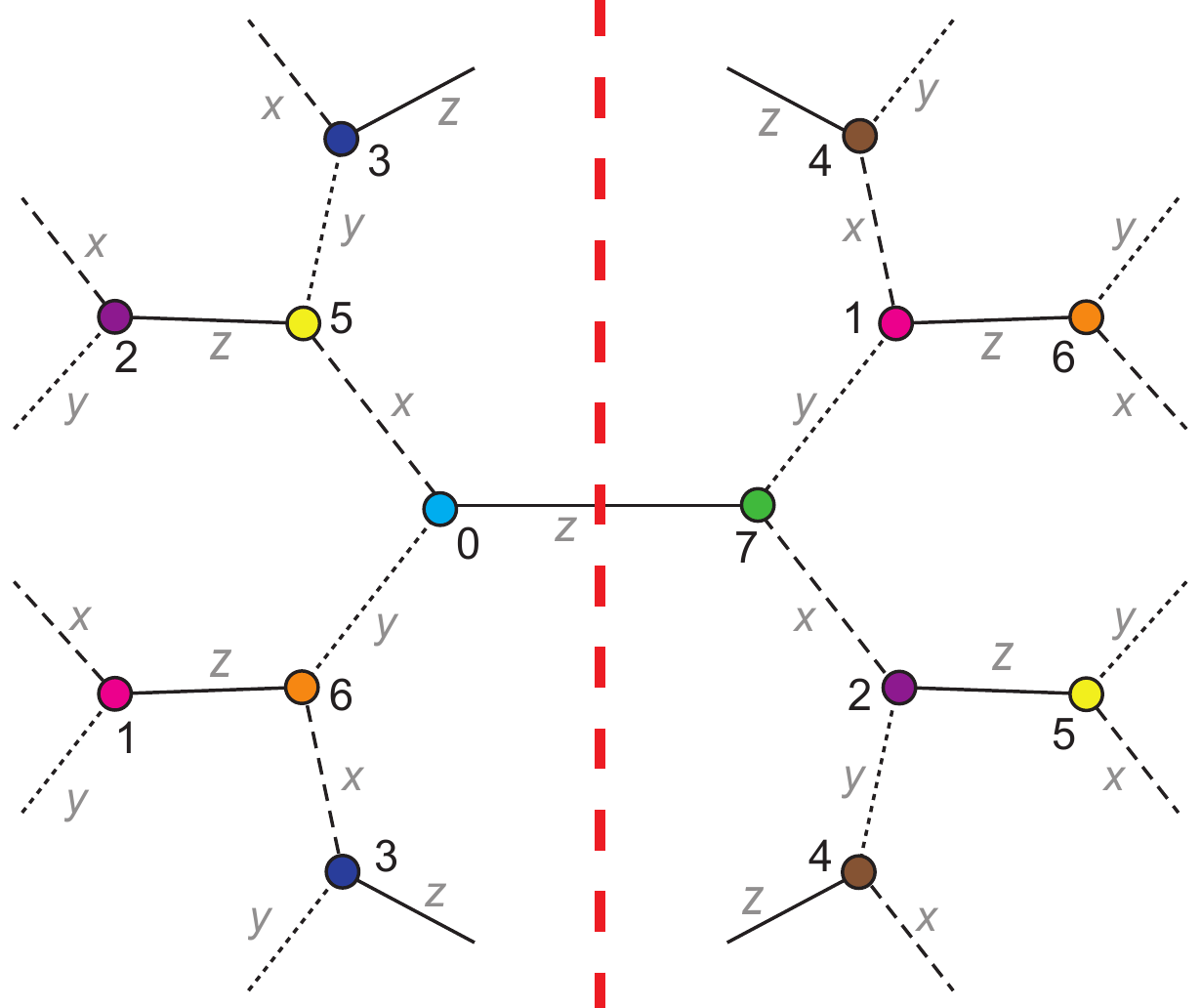}
\caption[]{{\bf Infinite-D $z{=}3$ tree with 8-site iTEBD cell.} 
Taking the $\ell{\rightarrow}\infty$ limit of the \hh lattice $\ell{=}10$ results in the $z{=}3$ tree, a lattice in infinite dimensions. 
Cutting any bond gives an entanglement bipartition (red dashed line), enabling entanglement-controlled computations with tensor network states. 
Bonds are labelled by Kitaev coupling; site labels show the unit cell used for the iTEBD computation. 
This 8-site unit cell (sites and bond Kitaev labels shown) used in the thermodynamic-limit iTEBD computation admits the Klein duality\cite{Khaliullin2005,Khaliullin2010,KHbhc}, and is thus expected to capture all the symmetry-breaking patterns in the phase diagram.}
\label{fig:tree}
\end{figure}

\subsection{Duality results for the magnetic phases}
\label{sec:magduality}
Even on the 3D lattices, definitive conclusions for the magnetically ordered phases can still be made due to a general feature, the \textit{Klein duality}, exhibited by Kitaev-Heisenberg models\cite{Khaliullin2005,Khaliullin2010,KHbhc}. The following discussion applies to any bipartitie lattice, including the tree lattice in infinite dimension, as well as all 3D harmonic honeycomb lattices. 
Since these lattices are bipartite, simple Neel antiferromagnetic order is the expected ground state for the Heisenberg antiferromagnet Hamiltonian. The Neel AF and FM orders map under the Klein duality to three dimensional generalizations of \textit{stripy} and \textit{zigzag} orders. Assuming that the unfrustrated Neel order is indeed the ground state for AF Heisenberg exchange (as may be verified by quantum Monte Carlo at the sign-problem-free SU(2) point), we conclude that all four of these magnetic phases must be present in the phase diagram. 

\subsection{Loop length as a control parameter}
\label{sec:loopcontrol}
To capture the full quantum phase diagram including the quantum spin liquid phases, we employ a limit inspired by the geometry in the hyperhoneycomb lattice.  Its shortest loops are the 10-site decagons. Treating this loop length $\ell=10$ as a large parameter and formally taking it to infinity, we find 
the loopless $\ell=\infty$ Cayley tree or Bethe lattice with $z=3$ connectivity in infinite dimensions. The tree lattice approximation  $\ell \rightarrow \infty$  enables a solution using entanglement-based methods originally developed for 1D systems, which rely on efficient representations of matrix or tensor product states (also known as projected entangled pair states PEPS). The key for such efficient representations is that entanglement is carried by bonds: cutting a single bond serves as an entanglement bipartition, and a singular value decomposition fully determines the entanglement spectrum which can be associated with this bond. 

This tree lattice is infinite dimensional in the sense that for finite trees with $N_s$ sites, a finite fraction of sites  $f_s\approx(z-2)/(z-1)$ is on the boundary. But note that this is an opposite limit of infinite dimensionality from the one commonly taken in mean field theories, which assume infinite connectivity $z\rightarrow \infty$: here we crucially fix $z=3$.  Entanglement based algorithms within our infinite-D approximation can work with the low coordination number $z=3$ and low spin $S=1/2$, capturing the associated strong quantum fluctuations. As discussed below, we employ an algorithm which studies the tree lattice directly in its thermodynamic limit, with no boundary sites, directly as an infinite system.

\subsection{Tensor networks on the tree lattice}
The large loop $\ell\rightarrow \infty$ limit of the hyperhoneycomb (or higher harmonic honeycomb) lattice, which yields the infinite-D Bethe tree lattice, admits a numerical solution of the gapped phases in the phase diagram. The key is that cutting a single bond gives an entanglement bipartition (as shown in Fig.~\ref{fig:tree}) with an entanglement spectrum which is associated with that bond. Hence gapped states can be represented efficiently as tensor product states (TPS, in other contexts also known as projected entangled pair states i.e. PEPS), and the full machinery of entanglement based algorithms can be used.
We choose to use a variant of the infinite system size time-evolving block decimation algorithm (iTEBD)\cite{Vidal2007}. The iTEBD algorithm has been previously used to study various Hamiltonians via tree tensor networks, with phase diagrams containing magnetic phases\cite{Nagaj2008,Nagy2012,Xiang2012}, nonmagnetic phases\cite{Su2013} and even a symmetry protected topological phase related to the AKLT Hamiltonian\cite{Depenbrock2013}. The iTEBD algorithm is especially useful here since it works directly in the thermodynamic limit (using iPEPS), avoiding the issues which plague finite trees.

Specifically, each update step in the algorithm, such as an imaginary time evolution step in iTEBD, must be followed by an operation which restores the state into a correctly normalized tensor product state.  This requires cutting the TPS into two parts and computing the entanglement spectrum across the cut, via a singular value (i.e. Schmidt) decomposition. These singular values are associated with the bond and placed between the adjacent site tensors when one contracts the TPS in order to measure observables. The tree lattice offers all these properties and hence entanglement based algorithms developed for 1D systems may be adapted to the tree\cite{Nagaj2008}. 

The iTEBD algorithm performs imaginary time evolution (i.e. soft projection to the ground state) within a restricted set of tensor product states, allowing it to find a good approximation to gapped periodic ground states with sufficiently local entanglement. Since it works on an infinite system, it always chooses one minimally entangled ground state, i.e. it can exhibit spontaneous symmetry breaking. To enable such symmetry breaking consistent with the expected magnetic ordering, we choose a unit cell with 8 site tensors and 12 bond (Schmidt) vectors as shown in Fig.~\ref{fig:tree}, employing 24 update cycles in each imaginary time evolution step.  On a technical note, we performed $2 \times 10^7$ singular value decompositions (SVDs) for each parameter point; to preserve normalized tensors during the imaginary time evolution, we intersperse evolution steps with zero imaginary time (i.e. pure SVD steps), as well as work with short time steps which are gradually reduced to $10^{-6}$ in inverse energy. The algorithm enables us to capture any periodic state consistent with our 8-site unit cell whose entanglement is sufficiently local, as is the case for the magnetically ordered phases we expect to find as well as for the gapped quantum spin liquids. 

The key parameter for TPS algorithms is the bond dimension $\chi$, serving as a cutoff on the number of entangled states.  The computational costs scale polynomially in $\chi$, but for computations on the tree the exponent is fairly high, with scaling of $\chi^6$. The Kitaev-Heisenberg model harbors additional computational complexity due to its lack of spin rotation symmetry, the large unit cell necessary to describe its magnetic phases and the emergent small energy scales in its quantum spin liquid phases.  Our results are roughly independent of $\chi$ for $\chi\geq 6$; we report data for computations using $\chi=12$, after verifying convergence through $\chi=6,8,10$. As we discuss below, the finite $\chi$ entanglement cutoff successfully collapses the degenerate ground-state manifold expected on the Bethe lattice into a single minimally entangled ground state, which is independent of these various values for $\chi$.  Hence we expect that finite (and perhaps not too large) $\chi$ is necessary for this mechanism which circumvents some of the issues which usually plague the Bethe lattice, but any $\chi$ within a large finite window will work well at enforcing a minimally entangled ground state.

\subsection{Definition of Hamiltonian parametrization}
The bond-anisotropic Kitaev-Heisenberg Hamiltonian, Eq.~\ref{eq:KH1}, involves one overall scale and and three free parameters.  In computing the quantum phase diagram via tensor product states, we will focus on two of these parameters.  The Kitaev exchange, generated by spin-orbit coupling, may be especially sensitive to the bond anisotropy; we therefore focus on the effects of bond anisotropy on the Kitaev term, leaving the study of the large-$\ell$ quantum phase diagram with Heisenberg term bond anisotropy for future work.  Note however that we have performed calculations on the full Hamiltonian Eq.~\ref{eq:KH1} in the neighborhood of the experimentally extracted parameter values shown in Fig.~\ref{fig:fits}, finding the magnetic Stripy-X/Y phase and a nearby phase boundary to the magnetic Stripy-Z phase. 

We shall now record the resulting two-parameter Hamiltonian, together with a few different useful parametrizations of its couplings, which we use to present various figures. In particular, we define  polar coordinates with $r=1-a$ and two different angle parameters, $\phi$ or the alternative $\theta$, corresponding to two differing conventions. The Hamiltonian parametrization is:
\begin{align}
\label{eq:KH2}
 &H = 
\sum_{\left< i j\right>} \left[ K_{\gamma_{i j}} S^{\gamma_{i j}}_i S^{\gamma_{i j}}_j + J  \vec{S}_i\cdot\vec{S}_j \right] 
 \\
 &K_{\gamma_{i j}} = K  *
\begin{cases}
  (1-a)  & \text{on}\ \ \gamma_{i j}=x,y  \  \text{bonds} \\
  (1+2a) & \text{on}\ \ \gamma_{i j}=z \quad \  \text{bonds}
\end{cases} 
\nonumber \\
 &K = 2 \sin(\phi), \quad  J = \cos(\phi); \quad \theta \equiv \pi/2 - \phi. \nonumber
\end{align}
The Kitaev-Heisenberg spin Hamiltonian, with the angular $\phi$ parametrization\cite{Khaliullin2012} relating the strengths of Kitaev and Heisenberg coupling, is extended with this symmetry-allowed anisotropy, parametrized by $-1/2 \leq a \leq 1$.

Let us here also note the extension of the Klein duality discussed in section~\ref{sec:magduality} above, to the case of nonzero anisotropy. 
Recall\cite{Khaliullin2012} that the Klein duality relates parameters by $\text{tan } \phi' = -(1+\text{tan } \phi)$ for the isotropic case $a=0$.  The transformation exposes a hidden ferromagnet even with anisotropy, at $\phi_\text{hidden FM}=-\text{arctan}[1/(1-a)]$. 
 Observe that the anisotropy reduces the symmetry at the hidden-FM point from SU(2) to U(1): the dual Hamiltonian is no longer Heisenberg but rather is an easy-axis ferromagnet. The key observation, that its ground state is an exact product state, remains unchanged.

For the pure Kitaev Hamiltonians, $a=1/4$ is the transition point between the gapless ($-1/2 \leq a\leq 1/4$) and gapped ($1/4<a \leq 1$) Z$_2$ spin liquid phases. In addition to the isotropic case $a=0$ we focus on a particular anisotropy value within the gapped QSL regime, $a=1/2$.  We sample other values of the anisotropy as well in order to generate the global phase diagram shown in Fig.~\ref{fig:radial_phases}. 

\subsection{Magnetically ordered phases}
Let us begin our analysis of the tensor product state computation by discussing the magnetic phases captured by the iTEBD algorithm on the tree lattice. 
We use a variety of measures to identify phases and the phase diagram.  Magnetically ordered phases can be captured directly by their magnetic order parameter, since the iTEBD always produces a single symmetry broken ground state. This analysis is shown in Fig.~\ref{fig:KH} for the isotropic model, and in the appendix in Fig.~\ref{fig:gapKHbig} for $a=1/2$ anisotropy. The four magnetic phases expected from the discussion in section~\ref{sec:magduality} above are observed. Phase transitions are also identifiable, as always, through the first and second derivatives of the energy. 
As a simple benchmark we have verified that the energy is always bounded from above by the energy of the expected classical product state and from below by the optimal energy for any given site in its surrounding cluster\cite{Anderson1951}, as may be seen in Fig.~\ref{fig:gapKHbig}.
Phase transitions are also signaled by peaks in the entanglement carried by the various bonds in the tensor product state, and finally of course the phases can be identified using the spin-spin correlation functions; these two measures are shown in Fig.~\ref{fig:KHcorrs}. We also verify that the entanglement correctly vanishes at the exact (hidden-)ferromagnet points.

The particular parameters of the direct first order phase transitions between the magnetic phases should be insensitive to dimensionality and loop length $\ell$ for sufficiently large $\ell$, since the quantum fluctuations on top of these classical phases need to propagate a distance of $\ell$ sites to distinguish one lattice from another.  The smallest value for $\ell$ we encounter is $\ell=6$, so quantum fluctuations in these magnetic orders must traverse at least six nearest-neighbor bonds to distinguish the honeycomb from the \shc or \hh lattices.  Hence we expect that the 2D honeycomb, 3D harmonic honeycomb and infinite-D tree lattices will exhibit similar magnetic transitions. Indeed the parameters we find for the tree lattice within iTEBD are essentially indistinguishable from those of the 2D honeycomb model\cite{Khaliullin2012}. As a function of anisotropy, the location of  magnetic transitions can also be compared to a classical mean field theory. We find similar behavior, with larger differences closer to the isotropic point; see Appendix~\ref{sec:magt} for details.

\begin{figure}[tbh]
\includegraphics[width=220 pt]{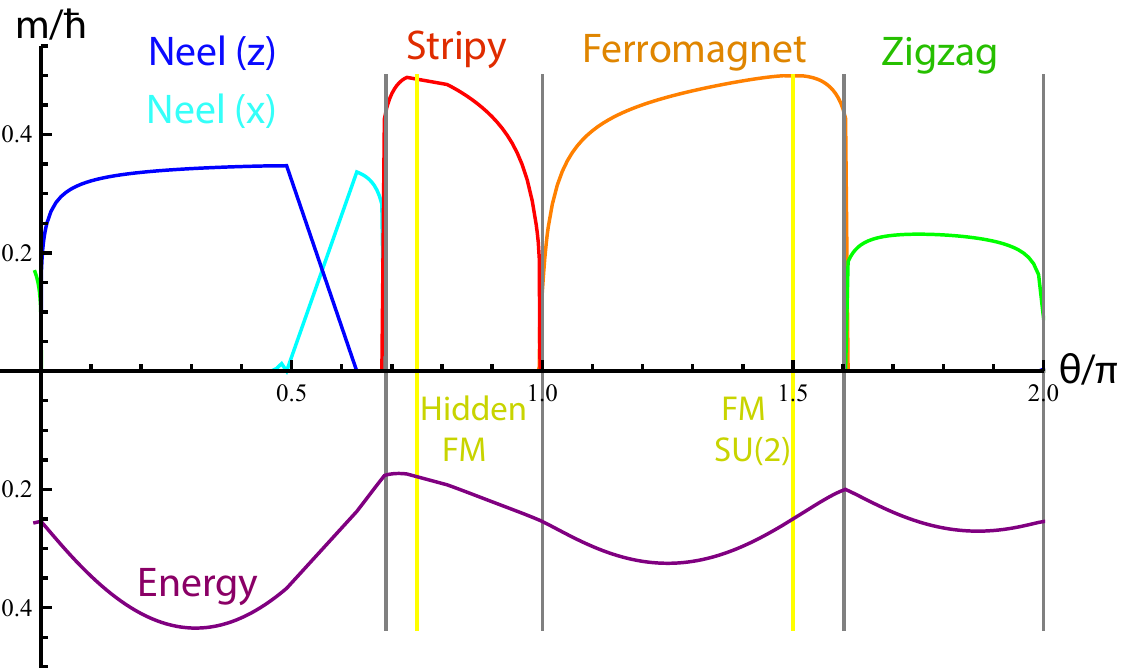}
\caption[]{{\bf Kitaev-Heisenberg isotropic phase diagram via iTEBD. } The magnetic order parameters of the four phases are directly observed by iTEBD, working in the thermodynamic limit. The stripy and FM phases surround an exact solution with saturated magnetic order parameter $m=\hbar/2$; Neel and zigzag phases exhibit a moment reduced by quantum fluctuations.  The energy per bond (purple) also provides the phase transitions, as well as benchmarking (see Fig.~\ref{fig:gapKHbig}). The QSL phases exist around the Kitaev points at $\theta/\pi=0,1$ but here are gapless and cannot be numerically captured with finite entanglement.}
\label{fig:KH}
\end{figure}

\subsection{The quantum spin liquid in a tree tensor network}
\begin{figure}[t]
\includegraphics[width=230 pt]{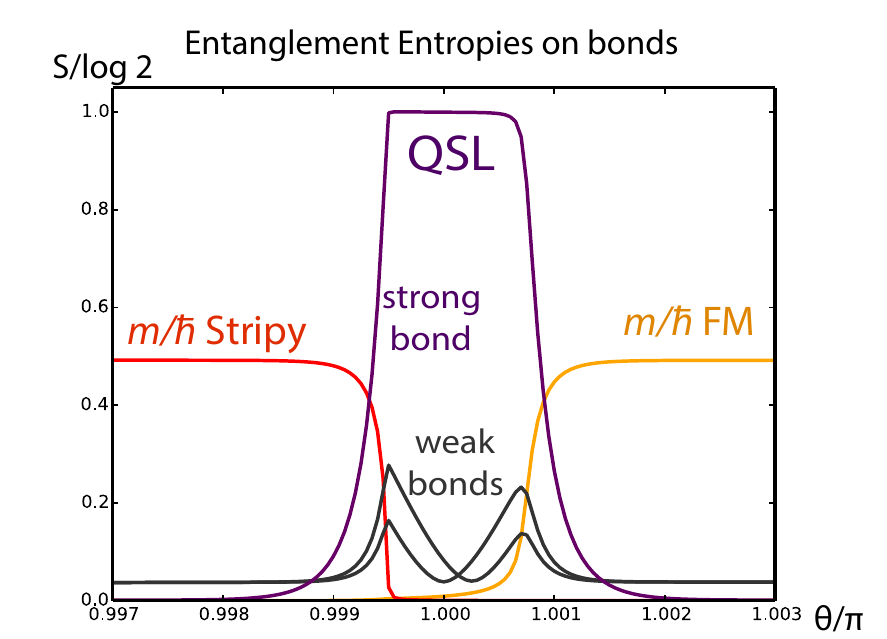}
\caption[]{  {\bf Gapped spin liquid at $K<0$ and surrounding magnetic phases, via $\chi=12$ iTEBD.} 
Sufficient bond anisotropy, here $a=1/2$, gaps the QSL fermion spectrum and enables a tensor product state representation. The QSL phase, here for $K <0$, is identified by the vanishing magnetic order parameters (here the stripy and ferromagnet) as well as by its entanglement entropies on the various bonds. The entanglement on the two weak bonds peaks at the transition (with slight spurious symmetry breaking), and that on the strong bond rises sharply in the QSL, matching the exact solution's entropy of Fig.~\ref{fig:anis}. }
\label{fig:gapKHqsl}
\end{figure}

Turning to the phase diagram of the QSL phases and their immediate surrounding, we first must restrict ourselves to the regime with sufficiently strong anisotropy so that the emergent majorana fermions are gapped, at $a>1/4$. The gapped spin liquids can be well approximated by the tensor product states we use. In Fig.~\ref{fig:gapKHqsl} we show the spin liquid phase for $K<0$ and the nearby stripy and ferromagnetic orders. The identity of the spin liquid is already suggested by its lack of magnetic order parameter;  phase transitions to this un-ordered phase are again seen in energy derivatives and as peaks in the entanglement entropies. The extent of the phase in this computation is small, covering about a tenth of a percent of the phase diagram, but nonzero; more importantly, the extent of the spin liquid is the same throughout the full range of bond dimensions we study, implying that its stability is a consequence of any finite entanglement cutoff.

Though its lack of conventional spin order is suggestive, the QSL phase completely lacks any order parameter and thus avoids a direct identification of the type in Fig.~\ref{fig:KH}.
The exact solution of the Kitaev model on the infinite-D tree allows us to uncover the unique fingerprints of the exact QSL, and use them to unequivocally identify the QSL phase within iTEBD. 
Each such set of fingerprints can be computed as a function of anisotropy for the pure Kitaev model across the entire gapped phase $1/4<a<1$. 

One obvious measure is the energy as a function of $a$ within the Majorana solution, for which we find good agreement as shown in Fig.~\ref{fig:aniscorr}; but energies are notoriously lousy fingerprints for spin liquid phases. We also compute the spin-spin correlators within the iTEBD and find that they match the correlators we compute within the exact solution, as shown again in Fig.~\ref{fig:aniscorr}. Correlation functions are a more robust measure, but are still grossly insufficient for fully characterizing the long ranged entangled QSL. 

\begin{figure}[t]
\includegraphics[width=200 pt]{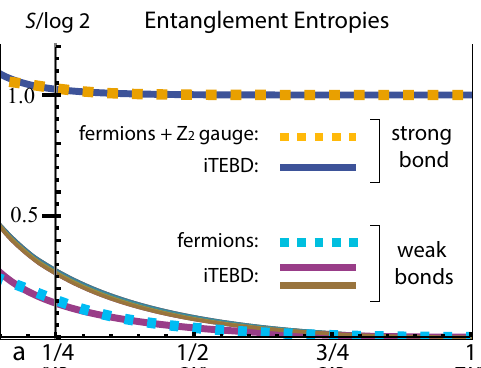}
\caption[]{{\bf Entanglement entropies as QSL fingerprints in iTEBD and exact majorana solutions.} 
Entanglement entropies from the exact solution with Majorana fermions and $\mathbb{Z}_2$ gauge fields (dotted lines) and from iTEBD computations (solid lines). The Kitaev bond strength anisotropy parameter $a$ is varied across the gapped phase $a>1/4$ for which the iTEBD algorithm can capture the quantum spin liquid.  The minimally entangled states of the QSL on the tree carry $\mathbb{Z}_2$ gauge sector entanglement only on strong bonds. }
\label{fig:anis}
\end{figure}

Instead, the most valuable set of fingerprints is furnished by the entanglement entropy carried by each bond. The entanglement spectrum is an inherent part of the tensor product state description and is easily accessible from the iTEBD. Spurious ``accidental'' symmetry breaking exhibited by the iTEBD ground state, caused by the large unit cell and the merely finite imaginary time evolution duration, complicates the bond entanglement entropies but still permits comparison with the entanglements computed in the exact solution. This comparison is shown in Fig.~\ref{fig:anis}, confirming that the iTEBD algorithm is indeed capturing the emergent Majorana fermions of the quantum spin liquid fractionalized phase in infinite dimensions.

Fig.~\ref{fig:anis} exhibits an important subtlety: the entanglement entropies from the exact solution match those from the iTEBD computation only if we assume that the gauge field sector contributes entanglement only on bonds set as strong by the anisotropy parameter.  To understand this key subtlety, we now turn to the study of the exact Kitaev $\mathbb{Z}_2$ quantum spin liquid on the loopless tree lattice, focusing first on the fermion sector followed by the more subtle $\mathbb{Z}_2$ gauge field sector.

\subsection{Majorana fermion entanglement}
The spectrum of Majorana fermions hopping on the infinite tree can be computed exactly\cite{Nagle1974} using recursion on propagators  (appendix \ref{appendixbethe}). However, we are mainly interested in the entanglement entropies associated with a bipartition, which we choose to compute on finite open trees. 
The spectrum of a finite tree adjacency matrix has an extensive number of zero modes, which 
may be counted for any finite tree by noting that the
number of bonds is $N_b=N_s-1$, reduced from the expected $N_s z/2$ by a fraction $f_b\approx (z-2)/z$; ie about $N_s/3$ of the eigenvalues are finite size boundary effects. 
For a site-centered tree they may be counted exactly 
 using Lieb's sublattice imbalance theorem\cite{Lieb1989} by observing the unbalanced occupation in the bipartite tree's  $A$ and $B$ sublattices, $|N_B-N_A|/N_s=(z-2)/z +1/N_s$. 
On a bond-centered tree, in addition to the identically zero boundary eigenvalues there is an isolated low-lying eigenvalue whose gap vanishes with increasing system size, which is also associated with the boundary. 
 We may thus take the bulk tree thermodynamic limit by discounting these boundary eigenvalues of finite tree adjacency matrices.

This finite tree thermodynamic limit, though convergent, may yield answers which are different from recursive computations directly on the infinite Bethe lattice for some physical quantities\cite{Nagle1974}\footnote{We thank Frank Pollmann for pointing out this subtlety.}. For example, the phase transition between the gapped and gapless phases computed by recursion equations for Green's functions (see appendix \ref{appendixbethe} for detail) find a phase boundary, as a function of hopping anisotropy, which is identical on the finite dimensional lattices but different on the Bethe lattice. However, we expect (and indeed show below) that total energy and the entanglement entropy in the thermodynamic limit of finite trees, with appropriate subtraction of the thermal entropy of the boundary described below, provide the correct thermodynamic limit for comparison with the iTEBD tensor network. 

Using the bulk fermion correlation function and the reduced correlator for a bipartition associated with cutting the central bond in a bond centered-tree, we first compute the entanglement spectrum and entropy of the bipartition, which resides on this central bond. A second approach for computing the entanglement entropy entails subtracting the $T=0$ thermal entropy of the finite open tree from the naively computed entanglement entropy of the bipartition, which again yields the entanglement entropy of the single bond cut without the $T=0$ thermal entropy of the boundary zero modes. See details in appendix \ref{appendixfermion}. The approaches agree, and thus are expected to yield the entanglement entropy contributed by the fermion sector of the exact quantum spin liquid.

\subsection{$\mathbb{Z}_2$ gauge theory on the loopless tree}
Entanglement is also contributed by the $\mathbb{Z}_2$ gauge sector of the tree Kitaev model;
in order to describe this contribution we shall now discuss the unusual subtleties which arise in a $\mathbb{Z}_2$ gauge theory on a loopless lattice. We begin by noting that the gauge theory is necessarily well defined even on the loopless tree lattice, since it arises from a well defined spin model.  For $N_s$ sites there are $N_s/2$ gauge-invariant sectors after projecting the gauge fields, which combine with the $N_s/2$ majorana fermion degree of freedom to give the $N_s$ doublet degrees of freedom for the lattice of $S=1/2$. In 2D the $N_s/2$ sectors are labeled by fluxes; in 3D, by magnetic field loops; and in infinite dimensions, they may be associated with $N_s/2$ particular infinite magnetic field lines extending across the (infinite) lattice. These field lines stretching across the system are intimately related to a more familiar set of infinite products of operators: in 2D topologically ordered phases, field lines can wind around the periodic boundary conditions. The resulting flux piercing the torus costs an energy which vanishes in the thermodynamic limit, and these operators generate the topological ground state degeneracy on the torus. On the tree lattice there is an extensive number of such operators, contributing an extensive ground state degeneracy $2^{N_s/2}$. 

This degeneracy may also be seen by counting conserved quantities associated with infinite products of local operators within the original quantum spin Hamiltonian.  Working either within the original spin model or within the gauge theory, we must count the number of such independent paths on the tree lattice. A moment's thought shows that independent paths may be counted as paths from a given boundary site to any other boundary site on a finite open tree. For the purposes of this counting the open tree may be compactified by identifying all boundary sites, in which case the strings again form conventional closed loops carrying a flux.  The counting gives exactly $N_{\text{boundary sites}}-1 = N_s/2$ operators, in agreement with the gauge field mechanism for $2^{N_s/2}$ degeneracy. Thus on the tree lattice within a full thermodynamic limit, the gauge theory collapses to an extensive degeneracy of $2^{N_s/2}$ states.

\subsection{Minimally entangled states of the $\mathbb{Z}_2$ gauge theory on the loopless tree}
\label{sec:Z2tree}
A $\mathbb{Z}_2$ gauge theory contributes $\log 2$ of entanglement for every two bonds in the entanglement cut, or $\log(2)/2$ entanglement per bond\cite{Yao2010}. Intuitively, the gauge field carries half of the information content of a physical gauge-invariant $\mathbb{Z}_2$ link variable. An additional global term of the topological entanglement entropy is generally expected to arise in the gauge theory, but does not appear on the tree lattice single-bond entanglement bipartitions: the only entanglement is that associated with the bond. Thus on the tree lattice we expect the single bond entanglement cut to carry  $\log(2)/2$ entanglement from the gauge sector, in addition to any fermionic entanglement. 

However, when comparing to the iTEBD result, we find that the iTEBD choice of ground state within the gauge theory's degenerate manifold effectively quenches the $\mathbb{Z}_2$ gauge sector entanglement on weak bonds, giving gauge sector entanglement only on strong bonds, which retain the gauge field entanglement $\log(2)/2$. This is reasonable since there are two $S=\log(2)/2$-carrying weak bonds per two sites, giving exactly the $\log 2$ value of entanglement associated with the twofold degeneracy also found per two sites. 
Thus for the iTEBD ground state on the tree, unlike for the unique ground state on the planar honeycomb, the entanglement entropy on various bonds is continuous in the Toric Code limit $a\rightarrow 1$, with weak bonds carrying vanishing entanglement like for the disjointed singlets Hamiltonian $a=1$.  The strong bonds carry entanglement of $\log(2)/2$ from the fermion sector plus $\log(2)/2$ from the gauge field sector, but for the weak bonds the fermionic entanglement vanishes and the gauge field entanglement is quenched by the minimally entangled superposition across flux sectors.

The finite bond dimension $\chi$ entanglement cutoff of the iTEBD algorithm is likely playing the key role here, collapsing the extensive degeneracy of the gauge theory on the tree into a particular minimally entangled state which is then chosen by iTEBD as the ground state. It will be interesting to explore whether this mechanism, of a ground state selected from an extensive degenerate manifold through a minimal-entanglement constraint, changes its role if the bond dimension is vastly increased.

Armed with the understanding of these subtleties, we thus find that aside from some spurious spontaneous symmetry breaking due to the infinite system size explored by the iTEBD algorithm, the entanglement entropy of the resulting iTEBD QSL ground state, as well as its energy and correlators, exhibit close agreement with these predictions of the exact QSL solution on a finite tree, as shown in Figs. \ref{fig:anis} and \ref{fig:aniscorr}. The TPS computation with the finite entanglement cutoff produces a minimally entangled state within the QSL manifold, elegantly bypassing artifacts due to the Bethe lattice lack of loops to successfully capture emergent Majorana fermions within the spin model at infinite dimensions.

\section{Conclusion}
In this work we have analyzed experimental data to motivate a magnetic Hamiltonian with large Kitaev exchanges, on the hyperhoneycomb and stripyhoneycomb lattices formed by Ir in  $\beta$- and $\gamma$-\li213.  Anisotropy in the strength of couplings between $z$ bonds and the $x,y$ bonds is expected from the crystal symmetries, and enables a fit to the experimental susceptibility measurements, requiring strong Kitaev exchange.  

We first focus on the pure-Kitaev models and discuss the exactly solvable 3D spin liquid, some of whose most interesting features are unique to three dimensionality. These features include the extended magnetic flux loop excitations as well as the existence of a finite temperature deconfined phase, neither of which can occur in the 2D honeycomb model. 
 
Describing the \li213 materials also requires some Heisenberg exchange, so we compute the quantum phase diagram as a function both of the additional Heisenberg exchange and of the coupling-strength bond anisotropy parameter.  Our approximation of choice is to study this system on the Bethe lattice, the tree with no closed loops. This is expected to capture the basic physics on the 3D harmonic honeycomb lattices, due to the long length of their shortest closed loop ($\ell=10$ for the \hh lattice or $\ell=6, 14$ for the \shc lattice). This large-$\ell$ approximation admits no analytical solution, but rather is numerically tractable via a class of entanglement-based algorithms. We use a TPS representation of the ground state, which is then determined using the iTEBD algorithm directly in the thermodynamic limit. 
Both the magnetically ordered phases as well as the gapped quantum spin liquid phases are obtained and positively identified using this technique. 

The exact 3D quantum spin liquid together with this large-$\ell$ approximation provide a controlled study of 3D fractionalization.  Although experimentally both of the 3D harmonic honeycomb Li$_2$IrO$_3$ polymorphs appear to be magnetically ordered\cite{Takagi2014, analytis2014}, the significant Kitaev couplings indicated by experiments are promising, and suggest future directions to realize 3D QSLs in these bulk solid state systems by  tuning magnetic interactions  via pressure or chemical composition.

\begin{acknowledgments}
We thank Christopher Henley, Masaki Oshikawa, Frank Pollmann, Ari Turner and Yuan-Ming Lu for inspiring discussions. I. K. thanks  Roderich Moessner, George Jackeli, Bela Bauer, Frank Pollmann, Olexei Motrunich, Kirill Shtengel and Duncan Haldane for useful comments when this work was presented at the SPORE13 workshop, MPIPKS, Dresden\cite{dresden}. This work was supported by the Director, Office of Science, Office of Basic Energy Sciences, Materials Sciences and Engineering Division, of the U.S. Department of Energy under Contract No. DE-AC02-05CH11231.  
\end{acknowledgments}

\section*{appendices}
\appendix

\section{Coordinates for the lattices}
\label{sec:vectors}
In this section, we define the 3D honeycomb-like lattices discussed in the paper.  For simplicity, throughout this paper we work with idealized symmetric versions of the true Ir lattices in the crystals.  

We  use the same parent orthorhombic coordinate system to describe both lattices.  This is the coordinate system defined by the following conventional orthorhombic crystallographic vectors:
\begin{equation}
\bm{a} = (2,2,0) , \quad  \bm{b} = (0,0,4), \quad \bm{c} = (6,-6,0). \end{equation}
In the equation above we have written the $a,b,c$ vectors in terms of a Cartesian (cubic orthonormal) $x,y,z$ coordinate system. The $\hat{x},\hat{y},\hat{z}$ lattice vectors in this coordinate system are defined as the vectors from an iridium atom to its neighboring oxygen atoms in the idealized cubic limit, with distance measured in units of the Ir-O distance. Nearest neighbors in the resulting Ir lattice are at distance $\sqrt{2}$.

For each lattice, we express its Bravais lattice vectors, as well as each of its sites of its unit cell, in terms of the $a,b,c$ axes.  A given vector or site, written as $(n_a,n_b,n_c)$, can be converted to the Cartesian coordinate system by the usual matrix transformation $(n_x,n_y,n_z) = n_a \vec{\bm{a}} + n_b \vec{\bm{b}}+ n_c \vec{\bm{c}}$.  For both of the lattices, the conventional crystallographic unit cell, containing 16 sites, is found by combining the primitive unit cell with the Bravais lattice vectors.

\subsection{Hyperhoneycomb lattice ($n=0$ harmonic honeycomb), space group $Fddd$ (\#70):}
Primitive unit cell (4 sites):
\begin{align}
\bigg(0, 0, 0\bigg); \ \left(0, 0, \frac{1}{6}\right); \ \left(\frac{1}{4}, \frac{-1}{4}, \frac{1}{4}\right); \ \left(\frac{1}{4}, \frac{-1}{4}, \frac{5}{12}\right)
\end{align}
This unit cell is formed by a single 16g Wyckoff orbit of $Fddd$, position $(1/8,1/8,z)$ with possible equivalent values of $z$ including $z=5/24$  (which shifts the unit cell above by $(1/8,1/8,1/24)$) and $z=17/24$ (with the same shift plus an additional $(1/2,0,0)$). 

Bravais lattice vectors (face centered orthorhombic):
\begin{equation} \left(\frac{1}{2}, \frac{1}{2}, 0 \right); \ \left(\frac{1}{2}, -\frac{1}{2},0 \right); \ \left(\frac{1}{2}, 0, \frac{1}{2} \right) . \end{equation}

\subsection{Stripyhoneycomb lattice ($n=1$ harmonic honeycomb), space group $Cccm$ (\#66):}
Primitive unit cell (8 sites):
\begin{align}
&\bigg(0, 0, 0\bigg); \ \left(0, 0, \frac{1}{6}\right); \ \left(\frac{1}{4}, \frac{-1}{4}, \frac{1}{4}\right); \ \left(\frac{1}{4}, \frac{-1}{4}, \frac{5}{12}\right); \nonumber\\
 &\left(0, 0, \frac{1}{2}\right); \ \left(0, 0, \frac{2}{3}\right); \ \left(\frac{1}{4}, \frac{1}{4}, \frac{3}{4}\right); \ \left(\frac{1}{4}, \frac{1}{4}, \frac{11}{12}\right) 
\end{align}
The sites $(0,0,1/6)$ and $(1/4,1/4,1/12)$  together represent the unit cell (shifted by $(0,0,1/6)$ from $(0,0,0)$) as the union of two distinct Wyckoff orbits, 8i with $z=1/6$ and 8k with $z=1/12$ ($Cccm$ origin choice 1). 

Bravais lattice vectors (base centered orthorhombic):
\begin{equation} \left(\frac{1}{2}, \frac{1}{2}, 0 \right); \ \left(\frac{1}{2}, -\frac{1}{2},0 \right); \ \bigg( 0, 0, 1\bigg) . \end{equation}

\section{Lattice tight-binding model and Majorana spectrum}
\label{sec:Mspectrum}
In the Kitaev spin liquid at its exactly solvable parameter point, the emergent Majorana Fermion hops within the nearest-neighbor tight binding model on the lattice. Its band structure (fixed to half filling) is given by the eigenvalues of the nearest-neighbor tight-binding matrix of the lattice.  

We now give these matrices for both lattices. 
For the \hh lattice, this matrix is
\begin{gather}
 \left(
\begin{array}{cccc}
 0 		& u^2 		& 0 		&\bu c_+ 	\\
 \bu^2	& 0 			& u c_- 	& 0 			\\
 0 		& \bu c_- 	& 0		& u^2		\\
 u c_+	& 0 			& \bu^2	& 0 			
\end{array} 
\right)
\end{gather}
We have used the following symbols to represent functions of wavevector $q$,
\[
 u=\frac{1}{\bu}=\exp\left(i \vec{q}{\cdot}\frac{\vec{c}}{12}\right),  \   c_\pm=2 \cos\left(\vec{q}{\cdot}\frac{(\vec{a}\pm \vec{b})}{4}\right) \]
To convert them to the Ir-O Cartesian axes, recall that $\frac{\vec{c}}{12} = \frac{\hat{x}-\hat{y}}{2} $ and $ \frac{(\vec{a}\pm \vec{b})}{4}=\frac{\hat{x}+\hat{y}}{2} \pm \hat{z}$.
For the \shc lattice, the unit cell has 8 sites and so we shall write an $8\times 8$ matrix.  By choosing an enlarged 8-site unit cell for the \hh lattice, we can represent the tight-binding matrices for both lattices in similar notation.  In the following matrix, the upper sign choice for the functions $c_\pm,c_\mp$ corresponds to the \shc lattice, while the lower sign choice corresponds to the \hh lattice with the enlarged unit cell.  These tight-binding matrices are
\begin{gather}
 \left(
\begin{array}{cccccccc}
 0 		& u^2 		& 0 		& 0 			& 0 			& 0 			& 0 			&\bu c_+ \\
 \bu^2	& 0 			& u c_- 	& 0 			& 0 			& 0 			& 0 			& 0 		\\
 0 		& \bu c_- 	& 0		& u^2		& 0 			& 0 			& 0 			& 0 		\\
 0 		& 0 			& \bu^2	& 0 			& u c_\mp 	& 0 			& 0 			& 0 		\\
 0 		& 0 			& 0 		& \bu c_\mp 	& 0 			& u^2 		& 0 			& 0		 \\	
 0 		& 0 			& 0 		& 0 			& \bu^2 	& 0 			& u c_\pm 	& 0 		\\
 0 		& 0 			& 0 		& 0 			& 0 			& \bu c_\pm 	& 0 			& u^2	 \\
 u c_+	& 0 			& 0 		& 0 			& 0 			& 0 			& \bu^2 	& 0
\end{array} 
\right) 
\end{gather}

The determinant of these matrices is the same for both lattices, simplifying to $\det = (1-2\cos(q{\cdot}c)S_{ab}+S_{ab}^2)$ with $S_{ab}{=}4(\cos(q{\cdot}a/2){+}\cos(q{\cdot}b/2))^2$. 
In this notation, it is evident that the zeros of the spectrum, found by setting the determinant to zero, are identical for both lattices and appear at the contour of momenta characterised by 
 the two
equations $\vec{q}{\cdot} \vec{c} = 0$  and $\cos \left( \vec{q} {\cdot}
\vec{a}/ 2 \right)+\cos\left( \vec{q} {\cdot} \vec{b}/ 2 \right) =
1/2$.  Note that the BZ for the 8-site unit cells is bounded by $\vec{q}{\cdot} \vec{c} = \pi$, $\vec{q}{\cdot} \vec{a} \pm \vec{q}{\cdot} \vec{b} = 2\pi$.

\section{Entanglement entropy from the Majorana fermions of the quantum spin liquid}
\label{appendixfermion}
At the exact QSL point we wish compute the entanglement entropy (and the energy) for the ground state on the tree, in order to compare this exact result to the iTEBD computation.  Within the gapped phase of the pure Kitaev (anisotropic) Hamiltonian, the system can be exactly mapped to a free majorana fermion problem with a gapped spectrum. We can thus compute quantities on finite trees independently of the iTEBD algorithm, within the majorana fermion mapping. Computing energies is straightforward and we find convergence to the thermodynamic limit using the boundary-eliminating procedure described above on trees with up to 9 layers.  To describe the entanglement entropy results, let us first recall the computation of entanglement spectrum and entropy for free fermion systems\cite{Peschel2003,Fidkowski2010,Qi2010,Vishwanath2010}. Operating on a bond-centered finite tree, we compute the correlation function by occupying half of the majorana spectrum. The reduced correlation function associated with a cut through the central bond is found by restricting the site indices of the correlator  to lie within one of the two partitions. Each eigenvalue $c_i$ of the reduced particle correlator also has an associated hole eigenvalue $1-c_i$. The entanglement entropy of the bipartition can be computed from the particle and hole correlators, with a factor of $1/2$ for majoranas, by $S_E = -(1/2) \sum_i [ c_i \log c_i + (1-c_i) \log (1-c_i)]$.

To eliminate tree finite size effects for computing the entanglement entropy in the fermion sector of the spin liquid, we use two approaches. In the first approach, we carefully determine which of eigenvalues of the open tree adjacency matrix are associated with the bulk, using the counting procedure described above, and keep only the eigenstates associated with these eigenvalues when computing the correlation function for the entanglement bipartition.  In the second approach, we subtract the $T=0$ thermal entropy of finite $L$-layered trees (with open boundary conditions) from the reduced density matrix entanglement entropy of each such tree under a bipartition through the center bond. 
This difference gives purely the entanglement entropy associated with the single bond cut, without the thermal entropy of the numerous zero modes of the boundary. 
We find agreement between the two approaches as the system size is increased (and the isolated boundary eigenvalue of the bond-centered tree vanishes).

\section{Anisotropic hopping on the infinite Bethe lattice}
\label{appendixbethe}
We compute the density of states on a Bethe lattice directly in the thermodynamic limit\cite{Rice1970}, where all sites are identical but each site has different hopping strengths $t_i$ ($i=1,...,z$) on the $z$ bonds connecting it to other sites.
Expressing the diagonal (onsite) Green's function in terms of a self energy,
\[ G(\omega) = \frac{1}{\omega(1-S(\omega))} \ , \quad S = \sum_{i=1}^{z} \sigma_i \]
where we suppress notational dependence on $\omega$; and where $\sigma_i$ is the self energy contributed from forward hopping starting from a $t_i$ hop. It obeys the following recursive system of questions:
\[ \sigma_i = \frac{t_i^2}{\omega^2} \frac{1}{1-\sum_{j\neq i}\sigma_j} \]
These may be rewritten as a set of quadratic equations, with implicit dependence only on $S$,
\[ \sigma_i^2 + (1-S)\sigma_i - t_i^2\omega^2 = 0\]
Solving this quadratic equation (the positive root is taken) and summing over $i$, we find a single equation for the self energy $S$.  We may then rewrite it directly as an implicit equation for the inverse Green's function $G^{-1}$,
\[ (z-2)G^{-1} + 2\omega = \sum_{i=1}^z \sqrt{(G^{-1})^2 + 4t_i^2} \]

The density of states $\rho$ is proportional to the imaginary part of $G$ (in this notation $\rho=-\text{Im } G/\pi$). The system is gapless here if there is a solution with nonzero DoS at zero energy. Writing $r=2\pi \rho(0)$, the equation to be solved is
\[ (z-2) = \sum_{i=1}^z \sqrt{1-t_i^2 r^2}\]
Let us analyze where a solution to this equation first appears. At $r=0$, the RHS is equal to $z$ and is greater than the LHS. The RHS decreases monotonically with $r$. However, $r$ takes values between $0$ and $r_{\rm max} = 1/\text{max}_i[t_i]$. The RHS takes its minimum value at $r_{\rm max}$. The phase boundary between the gapped and gapless phases occurs when this minimum value of the RHS is just barely equal to the LHS, i.e.
\[(z-2) = \sum_{i=1}^z \sqrt{1-\left(\frac{t_i}{t_{\rm max}}\right)^2}\ , \quad t_{\rm max} = \text{max}_i[t_i]\]

Let us consider this solution for the case when all hoppings $t_i=t$ are equal except one, $t_m$, which is larger than the rest. The phase boundary then occurs at
\[ t = \sqrt{2z-3} \frac{t_m}{(z-1)} \ ; \quad \ z=3\ \rightarrow \ a=\frac{2-\sqrt{3}}{2+2\sqrt{3}} \approx 0.05\]
In the loopless infinite $D=\infty$ Bethe lattice, the extent of the gapless phase is shrunk compared to its extent on the $D=2,3$ finite-dimensional lattices.  Indeed, the iTEBD computations, which are expected to break down for a gapless phase, are able to capture the gapped spin liquid characteristics down to $a\approx 0.15$, until they break down in a first order transition to full symmetry breaking.

\section{Harmonic honeycomb series}
In the notation for the $n$-harmonic honeycomb lattice,  the integer $n$ counts the number of hexagons forming the width of each fixed-orientation planar strip.
 Or equivalently, going along the direction of the special $c$ axis, the integer $n+1$ specifies the number of $z$-bonds between switches of the $x$,$y$ bonds orientation. 
Odd-$n$ lattices  posses a mirror plane perpendicular to the special axis, slicing through the midpoint of the (odd number of) hexagons; even-$n$ lattices possess no mirror reflections, only glide planes.  In this manuscript we focus on two lattices: the $n=1$ \shc lattice, space group \textit{Cccm}, recently synthesized\cite{analytis2014} as a polytype of Li$_2$IrO$_3$, with $\ell=6$ hexagon as well as $\ell=14$ sized minimal loops; and the $n=0$ hyperhoneycomb or \hh lattice, space group \textit{Fddd}, with $\ell=10$ decagon minimal loops. The two lattices are shown in Figures~\ref{fig:hhc1} and \ref{fig:hhc0} respectively.

The ``hyperhoneycomb'' terminology for the \hh lattice may be understood through the following connection to the hyperkagome lattice (also related to the hyperoctagon lattice\cite{Trebst2014}). 
Consider the 2D kagome and honeycomb lattices\footnote{this construction was suggested by Christopher Henley\cite{Henley2010,henleyprivate}.}. The kagome is the \textit{medial lattice} --- formed by connecting bond midpoints -- of the honeycomb lattice. This relation naturally suggests the existence of 3D honeycomb-like lattices which can be similarly associated with the 3D hyperkagome\cite{Takagi2007} lattice, the three dimensional lattice of corner-sharing triangles formed by iridium ions in Na$_4$Ir$_3$O$_8$. 
Indeed, the medial lattice of the \hh lattice has a graph (or, an adjacency matrix) which is, locally, identical to that of the hyperkagome: both feature corner-sharing triangles which combine to form $\ell=10$ decagon loops. These decagons arise from the $\ell=10$ minimal loops of the \hh lattice.

\section{Mean field for Stripy-X/Y order}
\label{sec:meanfield}
We briefly recall the self consistency equation for the mean field $S=1/2$ moment, 
$2|\vec{m_i}| =\tanh[|\vec{\widetilde{B_i}}|/(2 T) ]$ and $\hat{m_i} = \hat{\widetilde{B_i}}$, 
where  $\widetilde{B_i}$ is the mean field coupling to spin $S_i$. 
The spins develop a moment at a transition temperature $T_N = 2 E_{site}$, where $E_{site}$ is the classical energy per site in the ordered state (e.g. $-(1/4)(z/2)J$).
Above the transition temperature the mean field produces the Curie-Weiss law, $\chi^{r r} =  (g^r)^2  \mu_B^2 / \left(4 T + \sum_j J_{i_0 j}^{r r} \right)$, where  $\sum_j J_{i_0 j}^{r r}=zJ$ for nearest neighbor $J$. Experimentally relevant units may be restored by noting that $\mu_B = 0.672$ kelvin/tesla.  In the Stripy-X/Y order, the classical mean field $\vec{\widetilde{B_i}}$ takes the form
\begin{align}
\vec{\widetilde{B_i}} = g \mu_B \vec{B} -  ( & K_d m_i^x \hat{x} +K_d m_{n[i]}^y \hat{y} +K_c m_{n[i]}^z \hat{z} \\
										& +J_c \vec{m}_{n[i]}  +J_d \vec{m}_{n[i]} +J_d \vec{m_{i}}  ) \nonumber
\end{align}
and every site carries one of two magnetizations, $\vec{m_i}$ or $\vec{m}_{n[i]}$. 

Note that for the model Hamiltonians we consider, the principal axes of the susceptibility tensor are be $x,y,z$ rather than the crystallographic axes $a,b,c$.  Terms arising from the global symmetries of the crystal will likely change the principal axes to match the crystallographic ones. 
 To compare with experiment without adding any such additional terms, we measure the susceptibility tensor along the crystallographic axes as shown in Fig.~\ref{fig:fits}. The weakly anisotropic $g$-factors, experimentally determined at high temperature for each of the crystallographic axes, are then incorporated into each axis of $\chi$. We use the $g$-factors  $g^{x+y}=g^z=1.95$, $g^{x-y}=2.35$.  Note that the overall scale of the $g$-factors needed to fit the susceptibility, which was measured experimentally on a single crystal, carries an additional uncertainty associated with the uncertainty of estimating the number of \li213 formula units in the crystal.

\section{Comparison of magnetic transitions in iTEBD and mean field theory}
\label{sec:magt}
\begin{figure}[]
\includegraphics[width=245 pt]{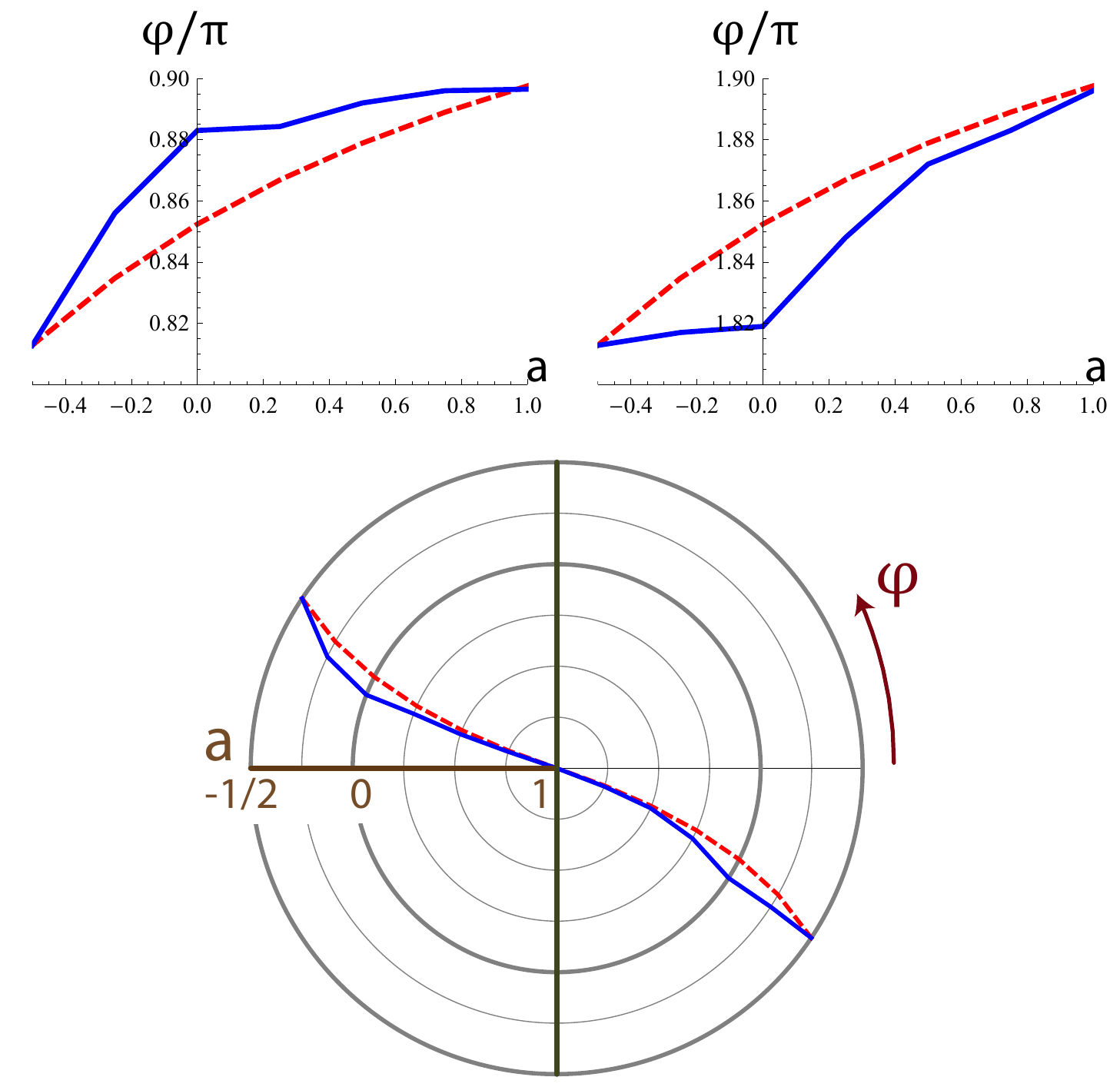}
\caption[]{{\bf Transitions between magnetic orders. } Comparison between iTEBD (solid blue line) and classical mean field theory (dashed red line), of the location in $\phi$ as a function of anisotropy $a$  of the first order transition between two adjacent magnetic orders.  Top left: transition between zigzag and FM phases. Top right: transition between stripy and Neel phases.  Bottom: transitions shown on the radial plot corresponding to Fig.~\ref{fig:radial_phases}.  }
\label{fig:magt}
\end{figure}

\begin{figure}[]
\includegraphics[width=110 pt]{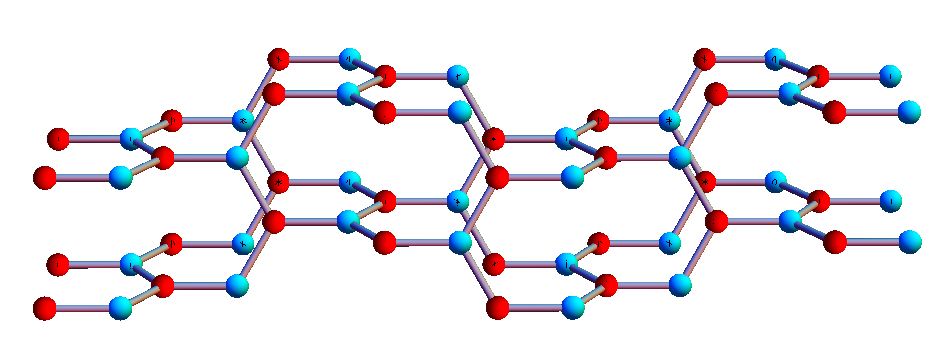}
\includegraphics[width=110 pt]{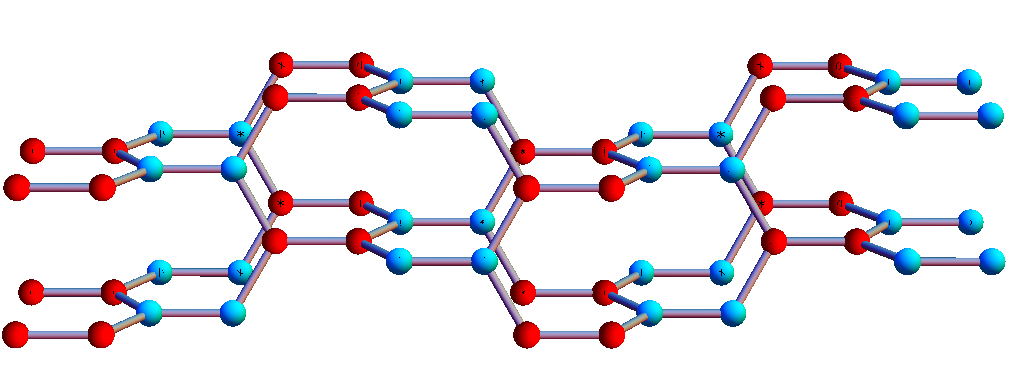}
\includegraphics[width=110 pt]{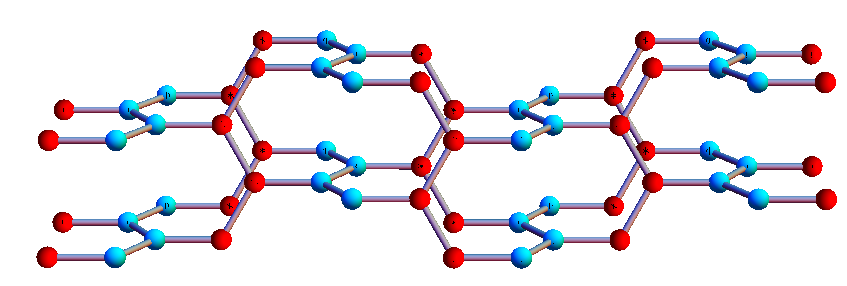}
\label{fig:magnetism}
\caption[]{  {\bf Magnetic orders on the hyperhoneycomb \hh lattice within the Kitaev-Heisenberg phase diagram.} Three magnetic configurations are shown: clockwise from top-left these are Neel, Stripy-Z and Zigzag-Z. 
Blue spheres denote up spins and red spheres denote down spins in these collinear antiferromagneic orders. Stripy-Z is dual to a $z$-oriented ferromagnet, Zigzag-Z is dual to a $z$-oriented Neel order. }
\end{figure}

Magnetic phases can be approximately described within a classical mean field theory. Such classical product states over sites, with no quantum fluctuation or entanglement, correspond to tensor product states with bond dimension $\chi=1$. On the Bethe lattice, we have captured the magnetically ordered phases using tensor product states with various $\chi$. We find that increasing $\chi$ to a value as low as $\chi=4$ is sufficient for capturing most of the quantum fluctuations near a first order transition between adjacent magnetic phases.  The location in parameter space of these transitions can be compared to the classical transition point.  Classically, the transition occurs at   $\phi=n \pi - \text{arccot}[2+a]$,  for anisotropy $a$, with $n=1$ for the zigzag-FM transition and $n=2$ fo the Stripy-Neel transition.  This comparison is shown in Fig.~\ref{fig:magt}.  For concreteness, we also draw sample magnetic configurations on the hyperhoneycomb lattice, shown in Fig.~\ref{fig:magnetism}. 

\section{Additional iTEBD results}
\begin{figure}[p]
\includegraphics[width=\columnwidth]{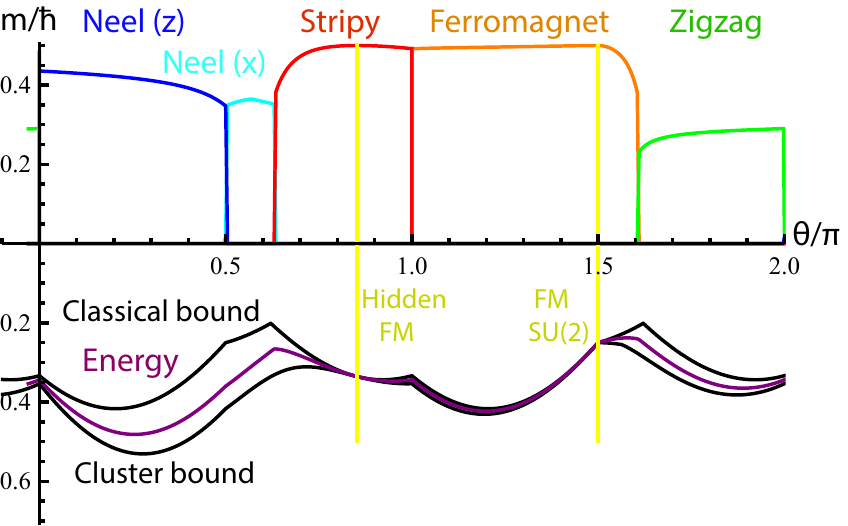}
\caption[]{{\bf Kitaev-Heisenberg magnetic phases with $a=1/2$ anisotropy. } See description for Fig.~\ref{fig:KH}. The energy measured in iTEBD is always found to be bounded from above by classical product states and from below by considering a maximally-entangled cluster, providing a check on the algorithm. The QSL phases here are gapped and shown in Figs.~\ref{fig:gapKHqsl} and \ref{fig:gapKHqsl2}, but their extent is not visible in this scale due to the strong anisotropy. }
\label{fig:gapKHbig}
\end{figure}

\begin{figure}[p]
\includegraphics[width=230 pt]{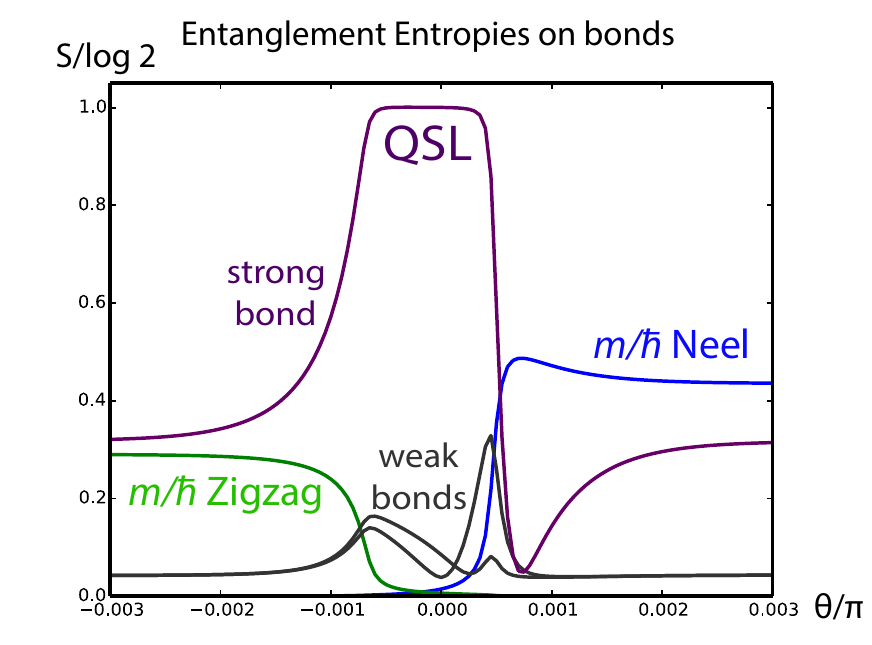}
\caption[]{   {\bf Gapped spin liquid for $K>0$  and surrounding magnetic phases within iTEBD.}
See the description in Fig.~\ref{fig:gapKHqsl}. Here we show the QSL at $a=1/2, K>0$, which competes with the zigzag and Neel orders. Different entropy curves occur here compared to the $K<0$ QSL since while both the FM and stripy orders are effectively ferromagnets with nearly saturated ordered moments, the Neel and zigzag phases involve substantial quantum fluctuations. }
\label{fig:gapKHqsl2}
\end{figure}

\begin{figure}[p]
\includegraphics[width=220 pt]{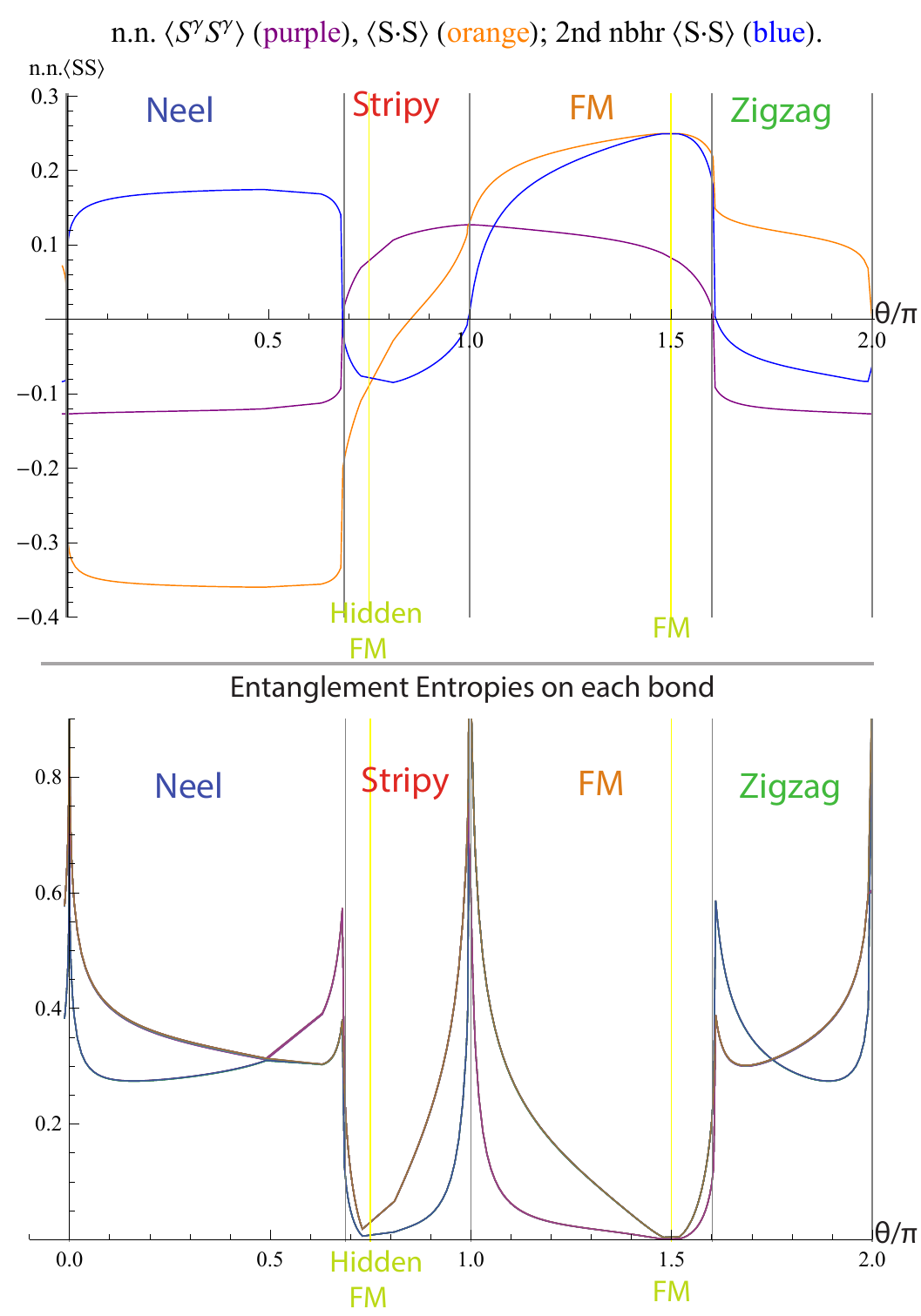}
\caption[]{ {\bf Isotropic $a=0$ Kitaev-Heisenberg  phase diagram via iTEBD: correlators and entanglement. } 
The four magnetically ordered phases can be identified by various measures in addition to their direct order parameters. These include signatures of the transitions in energy derivatives (not shown here), spin-spin correlators (top) and entanglement entropies on the various bonds in the unit cell (bottom). The entanglement entropies vanish at the exactly solvable points (shown by yellow lines) where the ground state, a (hidden) ferromagnet, is a simple product state.}
\label{fig:KHcorrs}
\end{figure}

\begin{figure}[p]
\includegraphics[width=200 pt]{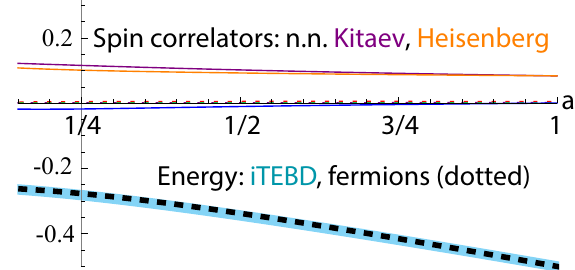}
\caption[]{{\bf Further benchmarks of iTEBD for the QSLs. } The iTEBD spin correlators match the expected result for the pure Kitaev model, vanishing except for nearest neighbor Kitaev-matched spins; all magnetic order parameters vanish (shown); and the iTEBD ground state energy per bond (cyan) matches the energy computed from the majorana fermion spectrum (black), in the gapped QSL phase. }
\label{fig:aniscorr}
\end{figure}

Here we present additional figures with results from the iTEBD computation, as described in the main text and in the figure captions. The results are shown in Figures  \ref{fig:gapKHbig}, \ref{fig:KHcorrs}, \ref{fig:gapKHqsl2}, and \ref{fig:aniscorr}.

\clearpage

\bibliography{IridatesCitations}

\end{document}